\documentclass[aps,pre,preprint,onecolumn,citeautoscript,superscriptaddress]{revtex4-1}  
\usepackage{amsmath,amssymb,bm} 
\usepackage{graphicx}  
\usepackage[tight]{subfigure}    
\usepackage{color} 
\usepackage{mathtools}
\usepackage[table]{xcolor}
\usepackage{multirow}
\usepackage[papersize={8.5in,11in}]{geometry}
\usepackage[colorlinks=true]{hyperref}  
\hypersetup{
    bookmarks=true,         % show bookmarks bar?
    unicode=false,          % non-Latin characters 
    pdftoolbar=true,        % show Acrobat
    pdfmenubar=true,        % show Acrobat 
    pdffitwindow=false,     % window fit to page when opened
    pdfstartview={FitH},    % fits the width of the page to the window
    pdftitle={My title},    % title
    pdfauthor={Author},     % author
    pdfsubject={Subject},   % subject of the document
    pdfcreator={Creator},   % creator of the document
    pdfproducer={Producer}, % producer of the document
    pdfkeywords={keyword1} {key2} {key3}, % list of keywords
    pdfnewwindow=true,      % links in new window
    colorlinks=true,       % false: boxed links; true: colored links
    linkcolor=magenta, %red,          % color of internal links (change box color with linkbordercolor)
    citecolor=blue,        % color of links to bibliography
    filecolor=magenta,      % color of file links
    urlcolor=cyan           % color of external links
} 

\newcommand{\beq}{\begin{equation}}
\newcommand{\eeq}{\end{equation}}
\newcommand{\nn}{\nonumber \\}
\newcommand{\eeqref}[1]{Eq.~(\ref{#1})}
\def\bea{\begin{eqnarray}}
\def\eea{\end{eqnarray}}
\begin{document}
\title{Spectrum of the Wilson-Fisher conformal field theory on the torus}

\author{Seth Whitsitt}
 \affiliation{Department of Physics, Harvard University, Cambridge, Massachusetts, 02138, USA}
 \author{Michael Schuler}
 \affiliation{Institut f\"ur Theoretische Physik, Universit\"at Innsbruck, A-6020 Innsbruck, Austria}
 \author{Louis-Paul Henry}
\affiliation{Institut f\"ur Theoretische Physik, Universit\"at Innsbruck, A-6020 Innsbruck, Austria}
\author{Andreas M. L\"auchli}
\affiliation{Institut f\"ur Theoretische Physik, Universit\"at Innsbruck, A-6020 Innsbruck, Austria}
 \author{Subir Sachdev}
 \affiliation{Department of Physics, Harvard University, Cambridge, Massachusetts, 02138, USA}
 \affiliation{Perimeter Institute for Theoretical Physics, Waterloo, Ontario N2L 2Y5, Canada}
 \begin{abstract}
We study the finite-size spectrum of the O($N$) symmetric Wilson-Fisher conformal field theory (CFT) on the 
$d=2$ spatial-dimension torus using the expansion in $\epsilon=3-d$. 
This is done by deriving a set of universal effective Hamiltonians describing fluctuations of the zero momentum modes. The effective Hamiltonians take the form of $N$-dimensional quantum anharmonic oscillators, which are shown to be strongly coupled at the critical point for small $\epsilon$. The low-energy spectrum is solved numerically for $N = 1,2,3,4$. Using exact diagonalization (ED), we also numerically study explicit lattice models known to be in the O($2$) and O($3$) universality class, obtaining estimates of the low-lying critical spectrum. The analytic and numerical results show excellent agreement and the critical low energy torus spectra are qualitatively different among the studied CFTs, identifying them as a useful fingerprint for detecting the universality class of a quantum critical point.
 \end{abstract}

\maketitle

\section{Introduction}

The identification of quantum critical behavior is an interesting problem in condensed matter and statistical mechanics. A major aspect of this is the emergence of universal low-energy behavior in the vicinity of a continuous quantum critical point, which is controlled by a CFT in the scaling limit.

In previous works, analysis of the torus spectrum has been used to study exotic deconfined quantum criticality, including transitions into phases with $\mathbb{Z}_2$ topological order \cite{HLSSW16,WS16} and conformal gauge theories coupled to fermionic matter \cite{TS16}.

In this paper we explore the finite-size energy spectrum of the Wilson-Fisher CFT, also known as the critical O($N$) model, in $(d+1)$ spacetime dimensions. One case where the structure of the spectrum is well-understood is when the system is on the $d$-dimensional sphere $S^d$. In this case, conformal invariance implies the \emph{state-operator correspondence}, which states that the energy spectrum takes the form $E_n = c\Delta_n/R$ where $c$ is the model-dependent speed of light, $R$ is the radius of the sphere, and the $\Delta_n$ are the scaling dimensions of the operators of the CFT in an infinite volume \cite{C84,C85}. 
These scaling dimensions are extremely constrained by conformal invariance, and the operator spectrum of many interesting CFTs has been mapped out using methods such as exact results available in two spacetime dimensions \cite{DMS97} and the conformal bootstrap \cite{El-Showk2012,El-Showk2014,Kos2014,Kos2014a,Kos2016}.
 
The state-operator correspondence has proven to be very useful in studying $(1+1)$-dimensional critical points, where numerically computing the spectrum on the circle is routinely done to accurately identify critical points \cite{Feiguin2007,Suwa2015}. However, in higher dimensions the curved geometry has proven to be difficult to implement accurately \cite{Alcaraz1987,Weigel2000,Deng2002,Brower2013,Brower2016}. 
In light of these difficulties, it seems natural to instead study the universal energy spectrum on flat geometries such as the torus, where the energy spectrum still takes the form $E_n = c \xi_n/L$ for some universal set of constants $\xi_n$ dependent on the shape of the torus. However, the structure of the torus spectrum is not simply related to the operator content, so one must use perturbative field theory.

In this paper we will use the $\epsilon$-expansion, where $\epsilon = 3 - d$, to compute the critical energy spectrum. We note that this method has already been applied by us to compute the universal spectrum for the Ising CFT \cite{HLSSW16}, but here we will generalize the result to general O($N$) models, and discuss how to compute the spectrum when the model is tuned away from criticality by a relevant perturbation. 
Two of us have also computed the spectrum on the torus directly in $(2+1)$-dimensions at leading order in the $1/N$ expansion \cite{WS16}; the correspondence between the $\epsilon$- and large-$N$ expansions is discussed in Appendix \ref{app:largen}. 

In this paper we will also numerically study the torus spectrum of explicit lattice models with critical points in the O(2) and O(3) universality classes using exact diagonalization (ED). We note that the numerical spectrum of the Ising CFT, O(1), has been discussed in detail in Ref.~\cite{HLSSW16}.  We will show that the numerical computations show excellent agreement with the $\epsilon$-expansion results. Beyond that, we will demonstrate that the critical low-energy torus spectra are intrinsically different among the distinct CFTs considered in this paper, characterizing the interpretation of the low-energy critical torus spectrum as a universal fingerprint of the underlying CFT and as a useful tool for investigating quantum critical points with diverse methods.

In Section \ref{sec:method} we will introduce our model and method, discussing how to treat fluctuations of the zero mode non-perturbatively using an effective Hamiltonian method. 
The main technical section of the paper is Section \ref{sec:calcs}, which details the structure of the effective Hamiltonians and how to compute them using perturbative quantum field theory, giving examples for several important special cases to demonstrate how the method works in general. 
In Section \ref{sec:num} we discuss how to numerically obtain the spectrum of the effective Hamiltonians in a few special cases, and give some results for the low-energy spectrum for the O($N$) models at $N = 2,3$. 
Finally, in Section \ref{sec:ED} we study several lattice models in the Wilson-Fisher universality class and numerically obtain their critical spectra using ED, and we compare numerical results to the analytic calculations.

\section{The method}
\label{sec:method}

The Wilson-Fisher CFT is described by the bare real-time Hamiltonian
\beq
H = \int d^d x \left( \frac{1}{2} \Pi_{\alpha}^2 + \frac{1}{2}\left( \nabla \phi_{\alpha} \right)^2 + \frac{s_0}{2} \phi_{\alpha}^2 + \frac{u}{4!} \phi_{\alpha}^4 + \Lambda \right) \label{L1}
\eeq
where the index $\alpha$ ranges from $1, ..., N$. We are using the notation $\phi_{\alpha}^2 \equiv \phi_{\alpha} \cdot \phi_{\alpha}$ and $\phi_{\alpha}^4 \equiv \left( \phi_{\alpha}^2 \right)^2$, so the model has full O($N$) symmetry. We suppress time-dependence, set the speed of light to unity, and note that fields satisfy the equal-time commutator $[\phi_{\alpha}(x),\Pi_{\beta}(x')] = i \delta_{\alpha\beta}\delta^d(x - x')$. We have included a bare ground state energy density $\Lambda$, which is needed to renormalize the ground state energy. The critical point is obtained by tuning $s_0 = s_c$, while $u$ approaches a fixed value $u^\ast$. This is a strongly-coupled theory for any finite $d < 3$ and $N$, but its universal properties can be computed as a power series in either $\epsilon = 3 - d$ or $1/N$.

\begin{figure}
\includegraphics[width=10cm]{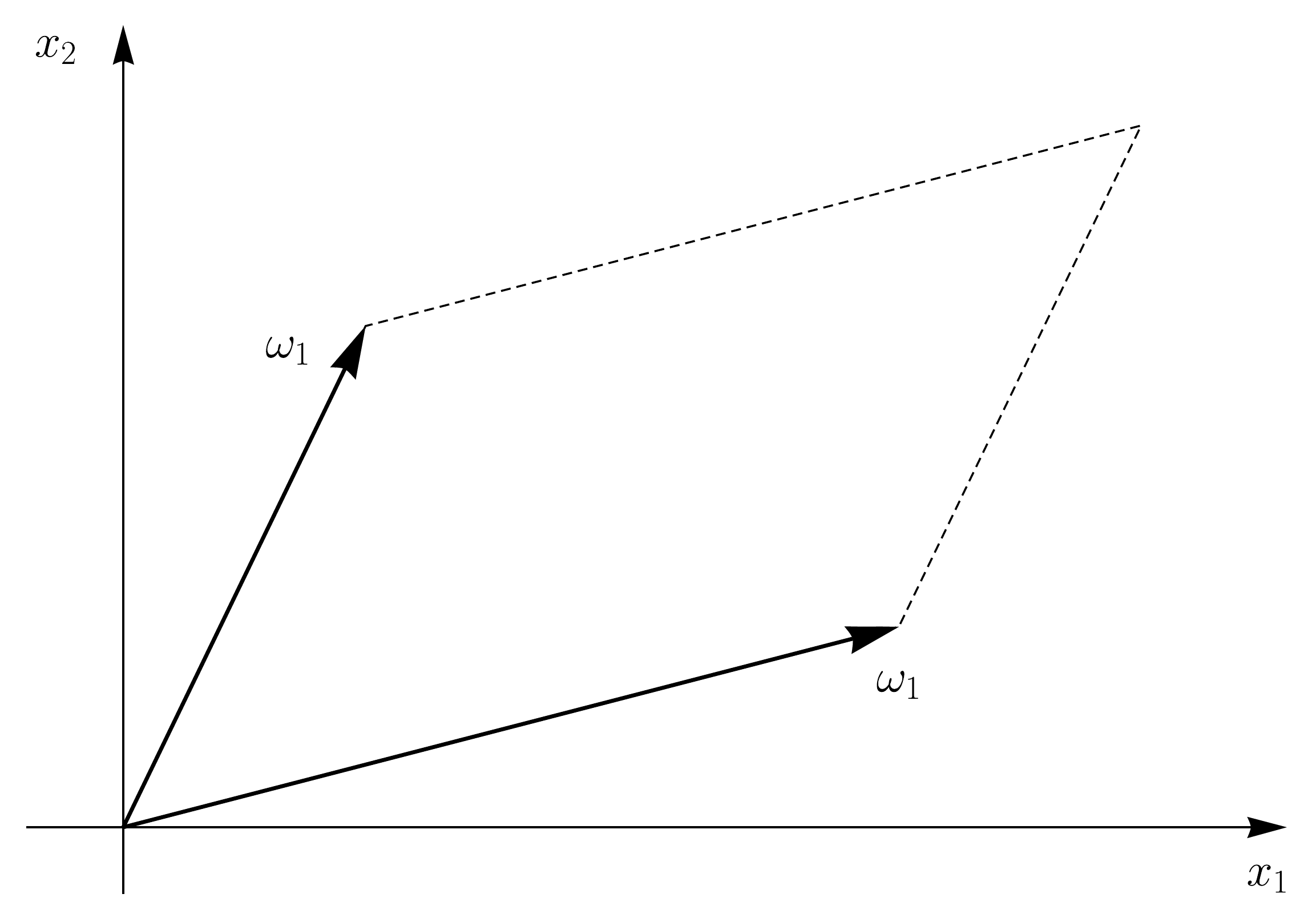}
\caption{The geometry of the torus on which our theory is defined. Here we have defined complex coordinates $x = x_1 + i x_2$, and we define the two complex periods $\omega_1$ and $\omega_2$. Then our geometry is the one pictured where the point $x$ and the point $x + n \omega_1 + m \omega_2$ are associated with each other for any $n,m \in \mathbb{Z}$.}
\label{fig:torus}
\end{figure}

We are interested in the finite-size spectrum of the above model on a spatial torus in $d = 2$, which is parametrized by complex coordinates $x = x_1 + ix_2$. We use the standard parametrization of the torus in terms of two complex periods, $\omega_1$ and $\omega_2$, and define the complex modular parameter $\tau \equiv \omega_2/\omega_1$ with real and imaginary parts denoted by $\tau = \tau_1 + i\tau_2$. Below, we will often give results in terms of the length scale $L \equiv | \omega_1|$. The area of the torus is given by $\mathcal{A} = \mathrm{Im}\left(\omega_2 \omega_1^\ast \right) = \tau_2 L^2$. This geometry is pictured in Fig.~\ref{fig:torus}. 
%In a previous work, we have calculated the spectrum to leading order in $1/N$ for $d=2$ \cite{WS16}. 

In this paper we calculate the spectrum in the $\epsilon$-expansion, which forces us to introduce extra dimensions \cite{WF72}. To this end, we consider $d/2$ copies of the desired two-dimensional spatial geometry, which retains the point-group and modular symmetries of the system while avoiding the introduction of any additional unphysical parameters. We then expand in $\epsilon$, and set $\epsilon = 1$ to obtain predictions for the $d=2$ system. 

As usual, we will need to eliminate the bare couplings in favor of finite renormalized couplings. Because renormalization is entirely due to short-distance divergences, the finite-volume theory on a manifold with no curvature must have identical renormalization constants to the infinite volume theory, since the only new length scale is large compared to the cutoff. Here we use a modified minimal subtraction scheme \cite{ZJ02}, where divergent poles in $\epsilon$ are subtracted with extra factors of
\beq
S_{d+1} = \frac{2}{\Gamma(d/2) \left( 4 \pi \right)^{d/2}} 
\eeq
attached. We introduce the renormalized coupling $g$ by
\beq
u = Z_4 \frac{\mu^{\epsilon} g}{S_{d+1}}
\eeq
In our calculations, we will always set $g$ to its fixed point value immediately after poles in $\epsilon$ have been subtracted. We also write 
\beq
s_0 = s_c + Z_2 s
\eeq 
where $s$ is a renormalized tuning parameter describing relevant perturbations across the critical point at $s=0$. 

We are also interested in the dependence of the ground state energy as a function of the geometry of the torus and the tuning $s$. This requires introducing the counterterm
\beq
\Lambda = Z_{\Lambda}\frac{s^2 Z_2^2 S_{d+1}}{\mu^{\epsilon}}
\eeq
which renders the energy density finite. In the above expressions, the $Z$ factors contain poles in $\epsilon$, and $\mu$ is an arbitrary energy scale. In principle one also needs to renormalize the fields $\phi_{\alpha}$, but these will not contribute to the leading order expressions so we ignore this here. For a review of these definitions and their relation to $L = \infty$ observables, see Appendix \ref{sec:infvol}.

The main technical feature of the $\epsilon$-expansion in a finite volume is the importance of the zero-momentum mode. Since the fields are gapless at the fixed point for $\epsilon \rightarrow 0$, the zero mode generates incurable infrared divergences in perturbation theory. These can be related to the failure of expanding around mean field theory, where the zero mode can have arbitrarily large fluctuations in the absence of interactions. This results in a continuum of zero mode excitations for the free field theory, whereas the interacting theory must have a gap and a discrete set of states, since the quartic term in the Hamiltonian will suppress fluctuations of the zero mode. Therefore, the free field theory with a zero mode is not the correct starting point in perturbation theory.

As was first realized by L{\"u}scher \cite{L82}, and further developed by others studying finite size effects in classical critical phenomena \cite{EBJZ85,RGJ85}, the solution is to separate the zero mode and treat it non-perturbatively. Since the finite momentum modes have an effective gap, they can be safely integrated out in a path integral approach, leading to an effective action for the zero momentum modes which must be treated exactly. Here we pursue a Hamiltonian approach rather than a path integral approach, but our method is the same in principle. We note that a similar real-time approach was used to study low-energy spectrum of Yang-Mills theory on the torus in Ref.~\cite{L83}.

We expand the fields and their conjugate momenta as
\bea
\phi_{\alpha}(x) &=& \mathcal{A}^{\frac{1-d}{4}}\varphi_{\alpha} + \frac{1}{\mathcal{A}^{d/4}} \sum_{k \neq 0} \frac{e^{i k \cdot x}}{\sqrt{2 \omega_{k}}}\left( b_{\alpha}(k) + b_{\alpha}^{\dagger}(-k) \right) \nn
\Pi_{\alpha}(x) &=& \mathcal{A}^{-\frac{d+1}{4}}  \pi_{\alpha} - \frac{i}{\mathcal{A}^{d/4}} \sum_{k \neq 0} \sqrt{\frac{\omega_{k}}{2}} e^{ik\cdot x}\left( b_{\alpha}(k) - b_{\alpha}^{\dagger}(-k) \right)
\eea
where $\omega_k = \sqrt{|k|^2 + s_0}$ and $k \cdot x = \mathrm{Re}(k x^\ast)$. The values of momentum summed over are determined by the shape of the torus, see Appendix \ref{dimreg}. Our expansion has been chosen such that the operators $\varphi_{\alpha}$, $\pi_{\alpha}$, and $b_{\alpha}(k)$ are dimensionless and have the commutation relations
\beq
[\varphi_{\alpha},\pi_{\beta}] = i \delta_{\alpha \beta}, \qquad [b_{\alpha}(k),b_{\beta}^{\dagger}(k')] = \delta_{\alpha \beta} \delta_{k k'}
\eeq
We now insert this expression into (\ref{L1}) and separate the Hamiltonian into a ``free'' and ``interaction'' part, where we insist that all zero-mode contributions are included in the interaction part.
\bea
H &=& H_0 + V, \nn
H_0 &=& \mathcal{E}_0 + \sum_{k \neq 0} \omega_{k} b^{\dagger}_{\alpha}(k)b_{\alpha}(k), \label{exact}
\eea
with the bare ground state energy
\beq
\mathcal{E}_0 = \mathcal{A}^{d/2} \Lambda + \frac{N}{2} \sum_{k \neq 0} \omega_k, \label{gseps1}
\eeq
and an interaction Hamiltonian,
\bea
V &=&  \frac{1}{\sqrt{\mathcal{A}}}\left\{ \frac{\pi^2}{2} + \frac{1}{2}\mathcal{A} s_0 \varphi^2 + \frac{u \mathcal{A}^{\epsilon/2}}{4!} \varphi^4 \right\} \nn
&+& \frac{1}{\sqrt{\mathcal{A}}}u \mathcal{A}^{\epsilon/2} \left( \frac{\delta_{\alpha \beta}}{12}\varphi^2 + \frac{1}{6} \varphi_{\alpha} \varphi_{\beta} \right) \sum_{\mathbf{k} \neq 0} \frac{1}{2 \mathcal{A}^{1/2} \omega_{k}} \chi_{\alpha}(-k)\chi_{\beta}(k) \nonumber \\
&+& \frac{1}{\sqrt{\mathcal{A}}} \frac{u \mathcal{A}^{\epsilon/2}}{6} \varphi_{\alpha} \sum_{k,k' \neq 0} \frac{1}{(8 \mathcal{A}^{3/2} \omega_{k}\omega_{k'}\omega_{k + k'})^{1/2}} \chi_{\alpha}(k) \chi_{\beta}(k') \chi_{\beta}(-k - k') \nn
&+& \frac{1}{\sqrt{\mathcal{A}}} \frac{u \mathcal{A}^{\epsilon/2}}{4!} \sum_{k,k',k'' \neq 0} \frac{1}{4 (\mathcal{A}^2 \omega_{k}\omega_{k'}\omega_{k''}\omega_{k + k' + k''})^{1/2}} \chi_{\alpha}(k) \chi_{\alpha}(k') \chi_{\beta}(k'') \chi_{\beta}(-k - k' - k'') 
\label{intH}
\eea
where $\chi_{\alpha} \equiv b_{\alpha}(k) + b_{\alpha}^{\dagger}(-k)$, and repeated latin indices are summed. We can now develop perturbation theory in $V$. The ground state energy (\ref{gseps1}) will be renormalized along with interactions in our final expressions and expanded to the appropriate order in $\epsilon$. 

Since the zero mode does not appear in the unperturbed Hamiltonian, the unperturbed eigenstates are given by
\beq
H_0 \Psi[\varphi_{\alpha}] | k,\alpha; k',\beta; \cdots \rangle = \left( \mathcal{E}_0 + \omega_k + \omega_{k'} + \cdots \right) \Psi[\varphi_{\alpha}] | k,\alpha; k',\beta; \cdots \rangle \label{psi0}
\eeq
Here, the energies are determined by the Fock states created by the $b_{\alpha}^{\dagger}$ operators, but states can be multiplied by \emph{arbitrary} normalizable functionals of the zero mode $\Psi[\varphi_{\alpha}]$. So the unperturbed states are infinitely degenerate, but this degeneracy is broken in perturbation theory.

We use a perturbation expansion developed by Bloch \cite{B58} which is well-suited to dealing with degeneracy. The method involves deriving an effective Hamiltonian within each degenerate subspace whose spectrum gives the splitting of that subspace. So we find an operator $H_{\mathrm{eff}}$ such that
\beq
H_{\mathrm{eff}} |\alpha_0 \rangle = E_{\alpha} |\alpha_0 \rangle
\eeq
where $|\alpha_0 \rangle$ are the set of unperturbed degenerate states, and 
\beq
E_{\alpha} = \epsilon_0 + \mathcal{O}(V)
\eeq
are the exact energies under the full interacting Hamiltonian, where $\epsilon_0$ is the unperturbed energy of the states $|\alpha_0 \rangle$. 

We review the derivation of Bloch's effective Hamiltonian in Appendix \ref{blochpert}. The main result needed is that the effective Hamiltonian at leading order is given by
\beq
H_{\mathrm{eff}} = \epsilon_0 P_0 + P_0 V P_0 + P_0 V \frac{1 - P_0}{\epsilon_0 - H_0} V P_0 + \mathcal{O}(V^3)
\label{eqn:effham}
\eeq
where $P_0$ is the projection operator onto the degenerate subspace $|\alpha_0 \rangle$. The calculation of the effective Hamiltonian will result in UV divergences due to summations over infinitely large momenta implicitly contained in Eq.~(\ref{eqn:effham}), so this is the step where the theory is renormalized.

From Eq.~(\ref{psi0}), we can infer the action of $H_{\mathrm{eff}}$ on our degenerate subspaces. An arbitrary state in a given degenerate manifold takes the form
\beq
\sum_{a = 1}^{M} \Psi_a[\varphi]  \left| \{ k_i \} \right\rangle_{a}
\eeq
where $M$ is the number of Fock states with the same energy. This degeneracy between inequivalent Fock states will be due to O($N$) symmetry or discrete rotation symmetry. The effective Hamiltonian will take the form of an $M\times M$ matrix which acts as
\beq
\sum_{b=1}^{M} H_{\mathrm{eff},ab} \ \Psi_b[\varphi]  | \{ k_i \} \rangle_{b} = E \ \Psi_a[\varphi]  | \{ k_i \} \rangle_{a}
\label{geneffham}
\eeq
This effective Hamiltonian will be a function of $\pi$ and $\varphi$. Due to the commutation relations, the action of $\pi$ on $\Psi_a[\varphi]$ is $\pi_{\alpha} = - i \frac{\partial}{\partial \varphi_{\alpha}}$, so this will be a set of $M$ coupled differential equations. In practice, symmetries of the interaction will allow us to consider smaller block-diagonal subspaces separately. For our model, the interaction conserves momentum and O($N$) rotations, which further constrains the form of $H_{\mathrm{eff}}$ and its eigenvectors.

\section{Calculation of the effective Hamiltonians}
\label{sec:calcs}
\subsection{Structure of the effective Hamiltonians}
\label{sec:structure}

Before proceeding with explicit calculations, we first describe the general structure of the effective Hamiltonian and its dependence on $\epsilon$, and discuss the perturbative spectrum of the Hamiltonian for small $\epsilon$. We give this discussion prior to explicit calculations because we will see that the perturbation theory is reordered in the scaling limit, leading to a modified expansion in fractional powers of $\epsilon$. We will find that the terms in $V$ do not contribute to the $\epsilon$-dependence of the spectrum that would be na\"ively inferred by Eq.~(\ref{intH}), so our analysis will aid us in correctly finding the leading contributions when we turn to explicit calculations. We will also highlight how the behavior of the spectrum changes depending on the magnitude of the tuning parameter $s$.

Here we will consider the effective Hamiltonian when the Fock state is non-degenerate ($M=1$ in Eq.~(\ref{geneffham})), since the analysis is similar in the general case. From Eq.~(\ref{eqn:effham}), the effective Hamiltonian at leading order takes the form
\bea
H_{\mathrm{eff}} &=& \mathcal{E}_0 + \frac{1}{\sqrt{\mathcal{A}}} \left[ K \frac{\pi^2}{2} + \frac{1}{2}R \varphi^2 + \frac{U}{4!} \varphi^4 \right] + \cdots \nn
K &=& 1 + \mathcal{O}(\epsilon^2) \nn
R &=& \mathcal{A} s + r_1 \epsilon + \mathcal{O}(\epsilon^2) \nn
U &=& \frac{48 \pi^2 \epsilon}{N+8} + \mathcal{O}(\epsilon^2)
\eea
where the ellipses will include additional operators which will not appear until $\mathcal{O}(\epsilon^2)$. The coefficient of each term will be a regular series in $\epsilon$. Here we write $r_1$ as the $\mathcal{O}(\epsilon)$ coefficient of the operator $\varphi_{\alpha}^2$. We have used the relations in Appendix \ref{sec:infvol} to set the quartic coupling to its fixed point. 

Since we are interested in the critical regime, we first consider the theory for $s = 0$. In this case, the coefficients of the quadratic and quartic terms are both $\mathcal{O}(\epsilon)$, but the structure of the spectrum can be made more clear by making the canonical transformation
\beq
\varphi \rightarrow \epsilon^{-1/6} \varphi, \qquad \pi \rightarrow \epsilon^{1/6} \pi
\label{contrans}
\eeq
after which the Hamiltonian is given by
\beq
H_{\mathrm{eff}} = \mathcal{E}_0 + \frac{\epsilon^{1/3}}{\sqrt{\mathcal{A}}} \left[ K \frac{\pi^2}{2} + \frac{1}{2}R \epsilon^{-2/3} \varphi^2 + \frac{U/\epsilon}{4!} \varphi^4 \right] + \cdots
\eeq
We see that when $s = 0$, the Hamiltonian takes the form
\beq
H_{\mathrm{eff}} = \mathcal{E}_0 + \frac{\epsilon^{1/3}}{\sqrt{\mathcal{A}}} h(\epsilon)
\eeq
where
\bea
h(\epsilon) &=& k \frac{\pi^2}{2} + \frac{1}{2}r \varphi^2 + u \varphi^4 \nn
k &=& 1 + \mathcal{O}(\epsilon^{2}) \nn
r &=& \epsilon^{1/3} \left(r_1 + \mathcal{O}(\epsilon) \right) \nn
u &=& \frac{48 \pi^2}{N+8} + \mathcal{O}(\epsilon)
\eea
We see that $h(0)$ is a pure quartic anharmonic oscillator whose spectrum gives the spectrum of $H_{\mathrm{eff}}$ to order $\epsilon^{1/3}$. Furthermore, the leading corrections are given by obtaining the spectrum of $h(\epsilon)$ in a perturbation series in $\epsilon^{1/3}$. Thus, the spectrum of the Wilson-Fisher fixed point at small $\epsilon$ has mapped to the spectrum of a quartic anharmonic oscillator in a strong-coupling expansion.

If we repeat the above analysis for $s \neq 0$, we find that the quadratic coefficient of the reduced Hamiltonian $h(\epsilon)$ is modified to
\beq
r = \frac{\mathcal{A} s}{\epsilon^{2/3}} + \epsilon^{1/3} \left( r_1 + \mathcal{O}(\epsilon) \right)
\eeq
For $\mathcal{A}s \gtrsim \epsilon^{2/3} $, our previous analysis no longer holds. The spectrum will be given by a weak-coupling expansion around a simple harmonic oscillator Hamiltonian provided $\mathcal{A} s \gg \epsilon$. This is sensible because our entire approach has been based on the gaplessness of the zero mode, whereas a nonzero $s$ contributes a gap. For large enough values of $\mathcal{A} s$, we do not need to separate the zero mode, and we could have done a normal expansion around Gaussian field theory, which is equivalent to the weak coupling expansion in the current effective Hamiltonian approach. The crossover from a weakly-coupled oscillator with a particle-like spectrum to a strongly-coupled oscillator signals the breakdown of particle-like excitations at the quantum critical point. It is interesting that this occurs already for arbitrarily small values of $\epsilon$, reflecting the importance of interactions in confining the zero mode.

In this paper it will be desirable to use the same strong-coupling expansion of $h(\varepsilon)$ at $s=0$ and $s \neq 0$, so we will always assume
\beq
\mathcal{A} s = \mathcal{O}(\epsilon)
\label{scalelimit}
\eeq
or smaller.

\begin{figure}
\centering
\includegraphics[width=16cm]{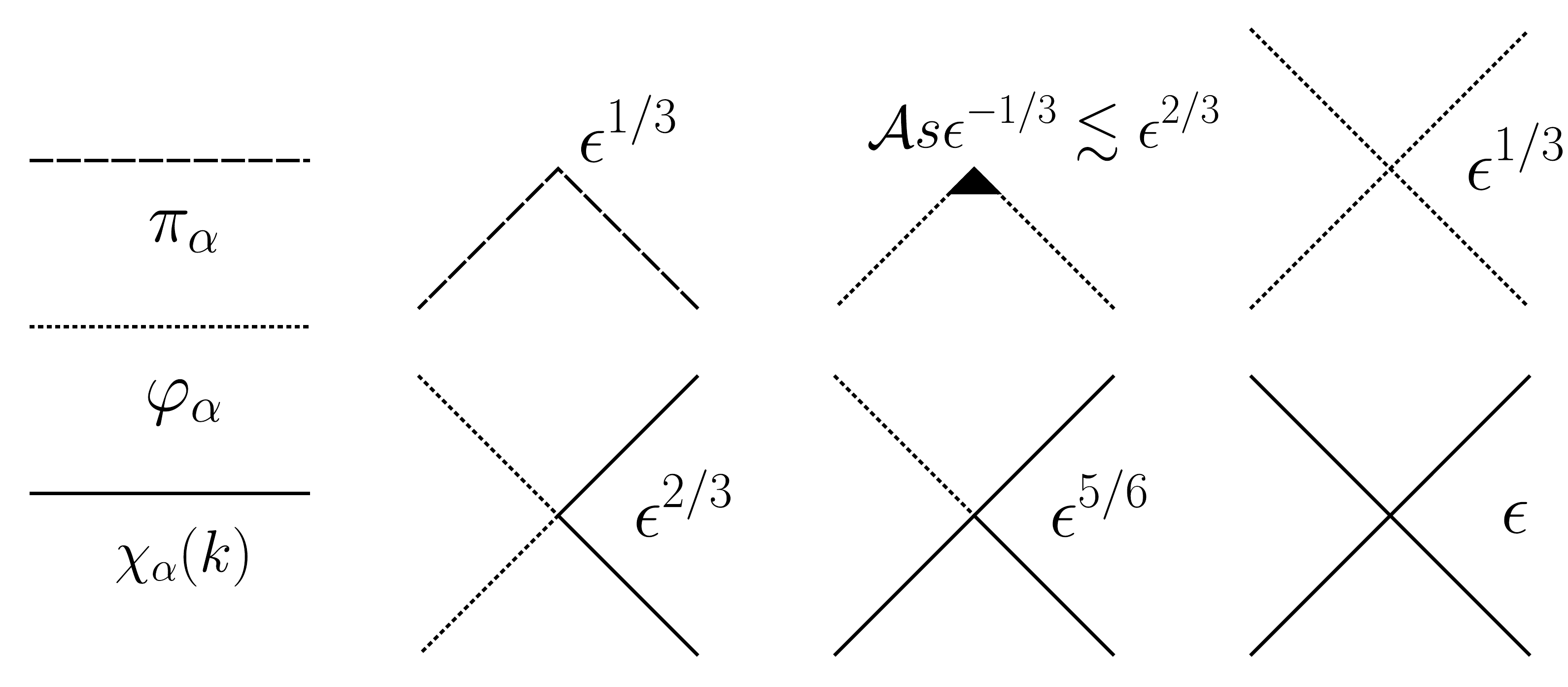}
\caption{(Left) Lines representing the insertion of the operators $\pi_{\alpha}$, $\varphi_{\alpha}$, and $\chi_{\alpha}(k) \equiv b_{\alpha}(k) + b_{\alpha}^{\dagger}(-k)$. (Right) The vertices of the interaction Hamiltonian coupling the zero modes to the finite momentum modes. The top three vertices represent the three terms on the first line of Eq.~(\ref{intH}), and appear in every effective Hamiltonian. The bottom three vertices represent the last three lines of Eq.~(\ref{intH}) respectively. As argued in Section \ref{sec:structure}, the labelled powers of $\epsilon$ refer to the order at which each vertex contributes to the spectrum.}
\label{fig:vertices}
\end{figure}

The reordering of the perturbative expansion requires a modified analysis of our perturbation theory. This can be most easily visualized by representing terms in the effective Hamiltonian diagrammatically, associating extra factors of $\epsilon^{-1/6}$ with factors of $\varphi$ anticipating the utility of the transformation in Eq.~(\ref{contrans}). With this in mind, we can schematically rewrite the interaction Hamiltonian Eq.~(\ref{intH}) after the canonical transformation to identify the individual terms with the correct powers of $\epsilon$:
\bea
V &=&  \frac{\epsilon^{1/3}}{\sqrt{\mathcal{A}}}\left\{ \frac{\pi^2}{2} + \frac{1}{2}\frac{\mathcal{A} s_0}{\epsilon^{2/3}} \varphi^2 + \frac{u \mathcal{A}^{\epsilon/2}}{\epsilon^{1/3} 4!} \varphi^4 \right\} + \frac{\epsilon^{2/3}}{\sqrt{\mathcal{A}}} \left( \frac{\delta_{\alpha \beta}}{12}\varphi^2 + \frac{1}{6} \varphi_{\alpha} \varphi_{\beta} \right) M^{(2)}_{\alpha \beta} \nonumber \\
&+& \frac{\epsilon^{5/6}}{\sqrt{\mathcal{A}}} \varphi_{\alpha} M^{(1)}_{\alpha} + \epsilon M^{(0)}
\label{modintH}
\eea
Here, the $M^{(n)}$ coefficients involve $n$ factors of finite-momentum modes. With this form of the interaction, we write down the vertices associated with $V$ in Fig. \ref{fig:vertices}. Then in calculating the effective Hamiltonian from Eq.~(\ref{eqn:effham}) we organize the $\epsilon$-expansion diagrammatically using these vertices. In practice it is easier to work directly with Eq.~(\ref{intH}) to compute the effective Hamiltonian, but the correct order of each term's contribution to the energy spectrum will be given by the $\epsilon$ coefficient pictured in Fig.~\ref{fig:vertices}. 

\subsection{Effective Hamiltonians for low-lying states}
\label{effhamcalcs}

In this section we will give the explicit derivation of the effective Hamiltonians for the lowest-lying states in the Fock spectrum. We will perform the calculation for increasingly complex cases, with each example having an added subtlety compared to the previous case, after which the general structure for the effective Hamiltonian splitting an arbitrary Fock state should follow.

\subsubsection{Fock vacuum}

We begin by considering the splitting of the Fock vacuum. This will give us the lowest-lying zero-momentum states, including the energy gap. The unperturbed eigenstate is
\beq
\Psi[\varphi] | 0 \rangle. \label{zpstate}
\eeq
Since $P_0 = |0\rangle \langle 0 |$, the effective Hamiltonian will be of the form
\beq
\mathcal{H}_{\mathrm{eff},k = 0} = h_{k = 0} |0 \rangle \langle 0 |.
\eeq
From Eq.~(\ref{geneffham}), the Schr\"odinger equation acting on the unperturbed subspace reduces to
\beq
h_{k=0}\Psi[\varphi] = E \Psi[\varphi]
\eeq
where, using Eq.~(\ref{eqn:effham}), the reduced Hamiltonian $h_{k=0}$ at one-loop is given by
\beq
h_{k = 0} = \mathcal{E}_0 + \langle 0 |V| 0 \rangle - \langle 0 | V \left( \frac{1 - | 0  \rangle \langle 0 |}{H_0 - \mathcal{E}_0} \right) V | 0 \rangle.
\label{eq:zeroeff}
\eeq

At this point we note that every term appearing in $h_{k=0}$ can be associated with a diagram. The three terms in this equation correspond to diagrams with zero, one, and two vertices respectively. Because the interaction $V$ conserves momentum, each vertex must also enforce momentum conservation. The expectation values and sums over $k$ implies one must contract all $\chi(k)$ propagators. Finally, the presence of the projector in the last term means we must contract the two vertices, preventing any disconnected diagrams from appearing.

\begin{figure}
\centering
\includegraphics[width=16cm]{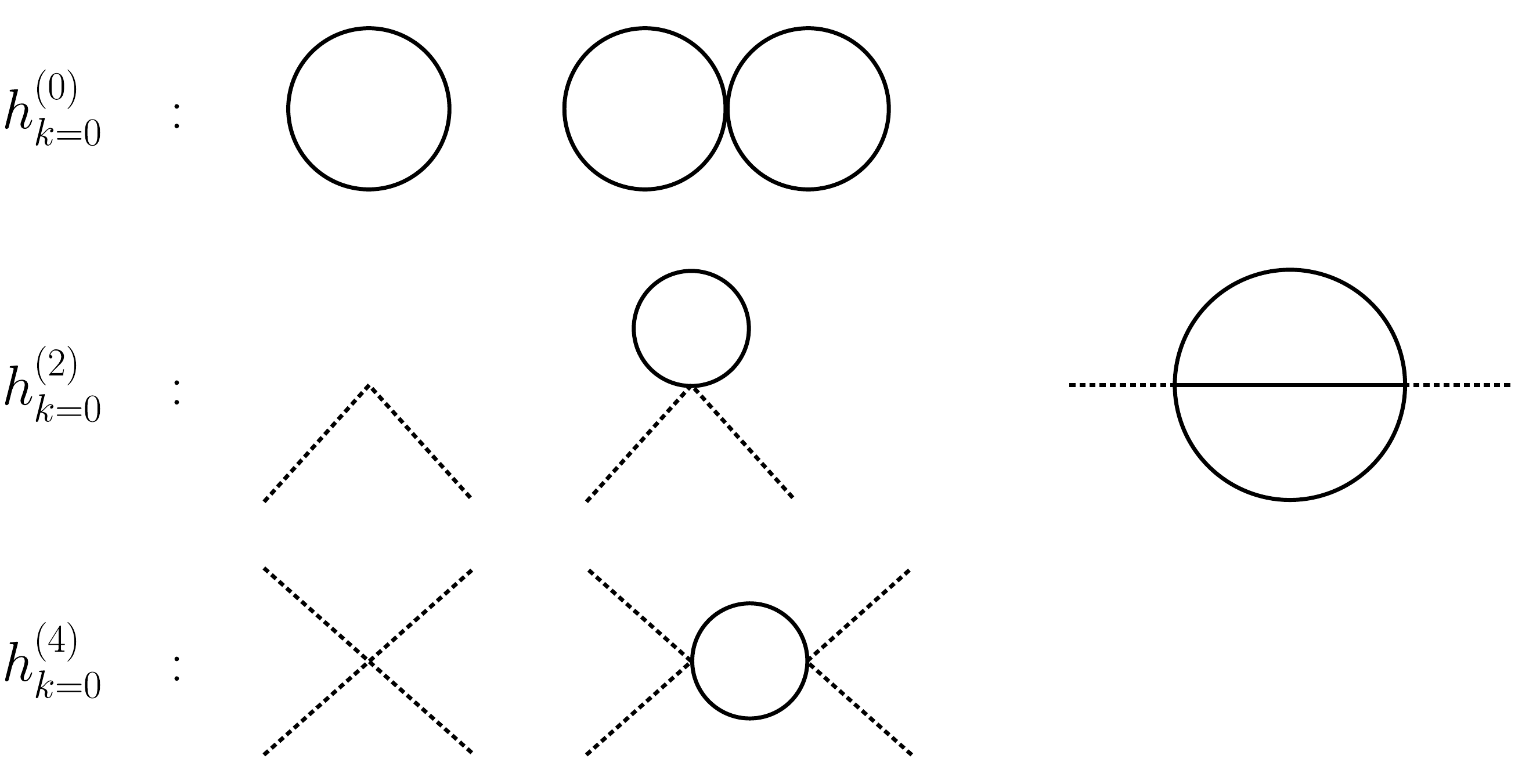}
\caption{(Left) The diagrams which contribute to the effective Hamiltonian at leading order. Each row is associated with a term written in Eq.~(\ref{eq:labels}). (Right) The leading two-loop contribution to the effective Hamiltonian, which we do not calculate.}
\label{fig:zerodiags}
\end{figure}

Writing the effective Hamiltonian as
\beq
h_{k=0} = \frac{1}{\sqrt{\mathcal{A}}} \frac{\pi_{\alpha}^2}{2} + h_{k=0}^{(0)} + h_{k=0}^{(2)} \varphi_{\alpha}^2 + h_{k=0}^{(4)} \varphi_{\alpha}^{4},
\label{eq:labels}
\eeq
we collect the one-loop terms which contribute to the effective Hamiltonian in Fig.~\ref{fig:zerodiags}. At $\mathcal{O}(\epsilon^{5/3})$ we encounter a nontrivial two-loop diagram, also pictured in Fig.~\ref{fig:zerodiags}, so we truncate the spectrum to order $\epsilon^{4/3}$. 

%We also need to justify truncating the effective Hamiltonian to third order in $V$. From Eq.~(\ref{eqn:effham2}), the third-order term is given by
%\beq
%h_{k = 0} \supset \langle 0 | V \left( \frac{1 - P_0}{\mathcal{E}_0 - H_0} \right)V \left(\frac{1 - P_0}{\mathcal{E}_0 - H_0} \right)V |0\rangle - \langle 0| V \left(\frac{1 - P_0}{\mathcal{E}_0 - H_0}\right)^2 V P_0 V |0\rangle.
%\eeq
%
%Using these rules, we see that the lowest-order two-loop diagram which contributes to the zero-mode part of the effective Hamiltonian is of order $\epsilon^{5/3}$, coming from a sunset diagram correcting the coefficient of $\varphi_{\alpha}^2$. Therefore we will keep all diagrams up to order $\epsilon^{4/3}$. Note that we include the two-loop contribution to the constant part of the Hamiltonian which is of order $\epsilon$. This diagram factorizes into the square of a one-loop diagram, so it is easily computed. Finally, we check that the lowest order diagrams with three vertices are of order $\epsilon^{5/3}$ (contributing the operators $\varphi_{\alpha}^6$ and $\pi_{\alpha}^2 \varphi_{\alpha}^4 + \mathrm{c.c.}$ to the Hamiltonian), which justifies our truncation. The diagrammatic representation of the statements in this paragraph is outlined in Figures \ref{}.

Using Eq.~$(\ref{intH})$, we obtain
\bea
h_{k=0}^{(0)} &=& \mathcal{A}^{(3-\epsilon)/2} \Lambda + \frac{N}{2} \sum_{k \neq 0} \sqrt{|k|^2 + s_0} + \frac{u \mathcal{A}^{\epsilon/2}}{\mathcal{A}^{3/2}} \frac{N(N+2)}{4(4!)} \left[ \sum_{k \neq 0} \frac{1}{\sqrt{|k|^2 + s_0}} \right]^2 \nn
h^{(2)}_{k=0} &=& \frac{\varphi^2}{2}\tau_2^{1/2} L s Z_2  + \frac{u \mathcal{A}^{\epsilon/2}}{2 \tau_2 L^2 } \left( \frac{\delta_{\alpha \beta}}{12}\varphi^2 + \frac{1}{6} \varphi_{\alpha} \varphi_{\beta} \right) \sum_{k \neq 0} \frac{\delta_{\alpha \beta}}{\sqrt{|k|^2 + s}} \nn
h^{(4)}_{k=0} &=& \frac{u \mathcal{A}^{\epsilon/2}}{\sqrt{\tau_2}L} \frac{\varphi^4}{4!} - \frac{\left( u \mathcal{A}^{\epsilon/2} \right)^2}{8 \tau_2^2 L^4} \left( \frac{\delta_{\alpha \beta}}{12}\varphi^2 + \frac{1}{6} \varphi_{\alpha} \varphi_{\beta} \right) \left( \frac{\delta_{\gamma \delta}}{12}\varphi^2 + \frac{1}{6} \varphi_{\gamma} \varphi_{\delta} \right) \sum_{k \neq 0} \frac{\delta_{\alpha \gamma}\delta_{\beta \delta} + \delta_{\alpha \delta}\delta_{\beta \gamma}}{\left( |k|^2 + s \right)^{3/2}}
\label{unrenorm}
\eea

At this point, we need to evaluate these infinite sums in $3 - \epsilon$ dimensions, and we need to renormalize the theory. These technical details are treated at length in the Appendices. In Appendix \ref{dimreg} it is shown how to evaluate infinite sums in arbitrary dimension, and in Appendix \ref{app:renorm} these three expressions are explicitly evaluated, and the cancellation of all divergences is demonstrated. We obtain the following effective Hamiltonian: 
\begin{gather}
h_{k = 0} = \mathcal{E}_{k = 0} + \frac{1}{\sqrt{\tau_2} L}\left( \frac{\pi_{\alpha}^2}{2} + \frac{R}{2} \varphi_{\alpha}^2 + \frac{U}{4!} \varphi_{\alpha}^4 \right) \nn
\mathcal{E}_{k = 0} \equiv \frac{ \pi N}{\tau_2 L} f^{(3 - \epsilon)}_{-1/2}(\tau,s,\mu) + \frac{1}{\sqrt{\tau_2} L} \frac{N(N+2)}{N+8}\frac{\epsilon}{8}\tau_2 f_{1/2}^{(3)}(\tau,s,\mu)^2 \nn
R \equiv \tau_2 L^2 s + 2 \pi \epsilon \left( \frac{N+2}{N+8} \right) \tau_2^{1/2} f_{1/2}^{(3)}(\tau,s,\mu) \nn
U \equiv \frac{48 \pi^2 \epsilon}{N+8} \left\{ 1 - \frac{\tau_2^{3/2} \epsilon}{4 \pi}f_{3/2}^{(3)}(\tau,s) + \frac{3\left(3N + 14 \right)}{\left( N+8 \right)^2}\epsilon \right\} 
\label{zeromodeham}
\end{gather}

Thus, the lowest states in the spectrum of the O($N$) Wilson-Fisher fixed point are given by solving the quantum mechanics problem of an isotropic, $N$-dimensional anharmonic oscillator. Here, the special functions $f^{(d)}_{\nu}(\tau,s,\mu)$ are given explicitly in Eqns.~(\ref{f-12}-\ref{f32}), and the function $f^{(3 - \epsilon)}_{-1/2}(\tau,s,\mu)$ should be expanded to first order in $\epsilon$. 

The functions $f^{(d)}_{-1/2}$ and $f^{(d)}_{1/2}$ depend on the renormalization scale $\mu$, which can be eliminated by applying renormalization conditions on the $s\neq0$ ground state energy and the energy gap. In this paper we will not eliminate $\mu$, since the infinite volume quantities are non-analytic around the critical point. However, our assumption that $s \sim \mathcal{O}(\epsilon)$ allows us to set $s$ to zero in many of the terms. We furthermore note that at the critical point, $s=0$, the $\mu$-dependence drops out and the spectrum is a universal function of $\tau$, $N$, and $\epsilon$. The coefficients of the Hamiltonian are modular invariant at $s=0$, which follows from Eq.~(\ref{fprops}) and the modular invariance of $\mathcal{A} = \tau_2 L^2$.

After performing a canonical transformation similar to Eq.~(\ref{contrans}), this Hamiltonian takes the form
\beq
h_{k = 0} = \mathcal{E}_{k = 0} + \frac{1}{\sqrt{\tau_2} L} \left( \frac{U}{4!} \right)^{1/3} \left( \frac{\pi_{\alpha}^2}{2} + \frac{RU^{-2/3}}{2} \varphi_{\alpha}^2 + \varphi_{\alpha}^4 \right)
\label{reducham}
\eeq
We will primarily work with this form of the Hamiltonian in Section \ref{sec:num}. 

\subsubsection{Single particle Fock states}

We now consider the splitting of the single particle state
\beq
\Psi_{\alpha}[\varphi] | k, \alpha \rangle \label{spstate}
\eeq
where we assume there are no multi-particle Fock states with the same momentum and energy, so we only need to consider an $N$-fold degenerate manifold. This assumption should hold for the smallest values of $|k|$. The effective Hamiltonian can be written as a matrix equation acting on the vector of functions $\Psi_{\alpha}$:
\beq
\sum_{\beta = 1}^N h_{k,\alpha \beta}\Psi_{\beta}[\varphi] = E \Psi_{\alpha}[\varphi]
\eeq
where the effective Hamiltonian $h_{k,\alpha \beta}$ can be represented by an $N \times N$ matrix whose components are
\beq
h_{k, \alpha \beta} = \left( \sqrt{k^2 + s_0^2} + \mathcal{E}_0 \right) \delta_{\alpha \beta} + \langle k, \alpha |V| k, \beta \rangle - \langle k, \alpha | V \left( \frac{1 - \sum_{\alpha}| k,\alpha  \rangle \langle k,\alpha |}{H_0 - \sqrt{|k|^2 + s} -\mathcal{E}_0} \right) V | k, \beta \rangle
\label{eq:singeff}
\eeq
and $|k,\alpha \rangle = b^{\dagger}_{\alpha}|0\rangle$.

\begin{figure}
\centering
\includegraphics[width=16cm]{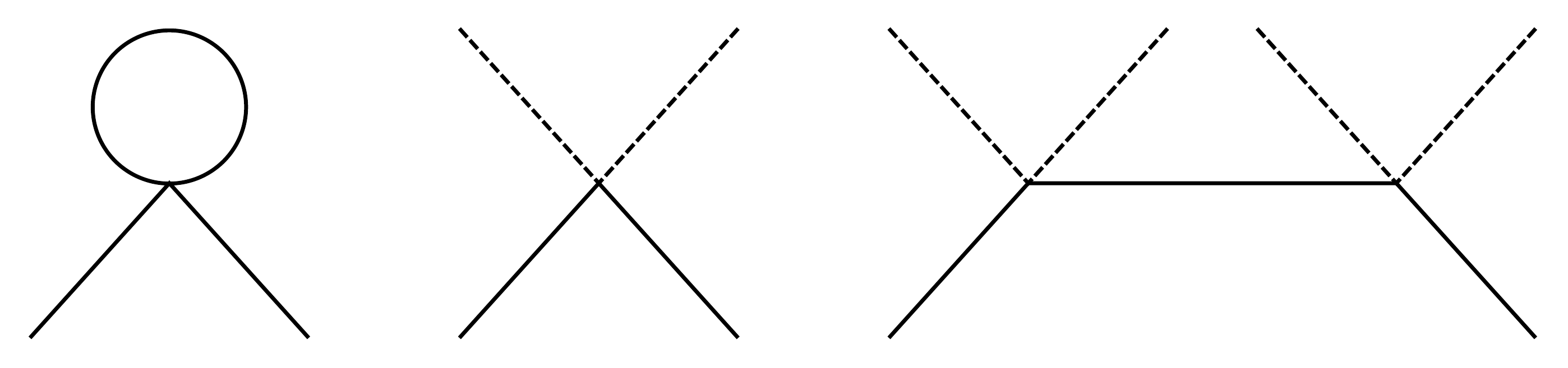}
\caption{Diagrams which contribute new terms to the effective Hamiltonian for the splitting of a single Fock state.}
\label{fig:singdiags}
\end{figure}

The terms in this equation can also be given a diagrammatic representation: one can consider all diagrams with two external finite-momentum lines carrying momentum $k$. However, the external momenta do not need to be contracted with the vertices in any of the terms. As a consequence, we obtain a term
\beq
h_{k,\alpha \beta} \supset h_{k = 0} \delta_{\alpha \beta}
\eeq
simply by taking the diagrams in Fig.~\ref{fig:zerodiags} and drawing a disconnected solid line in them. In addition to these, we will get new contributions due to the extra diagrams pictured in Figure \ref{fig:singdiags}. Only one of these diagrams contains a loop, so there will only be a single new divergence which is cancelled by the mass renormalization of $s_0$ in the first term of Eq.~(\ref{eq:singeff}). 

An explicit calculation similar to the one done for the Fock vacuum results in
\begin{gather}
h_{k,\alpha \beta} = \left( \mathcal{E}_k + h_{k = 0} \right) \delta_{\alpha \beta} + \frac{1}{\sqrt{\tau_2} L} \left( \frac{R_k}{2} \left( \varphi_{\eta}^2 \delta_{\alpha \beta} + 2\varphi_{\alpha} \varphi_{\beta}  \right) + \frac{U_k}{4!} \left( \delta_{\alpha \beta} \varphi_{\eta}^4 + \varphi_{\eta}^2 \varphi_{\alpha} \varphi_{\beta} \right) \right) \nn
\mathcal{E}_k \equiv \sqrt{|k|^2 + s} + \frac{1}{L} \left( \frac{N+2}{N+8} \right) \frac{\pi \epsilon}{\sqrt{|k|^2 + s}} f_{1/2}^{(3)}(s,\tau,\mu) \nn
R_k \equiv \frac{1}{\tau_2L^2}\frac{8 \pi^2 \epsilon}{(N+8)} \frac{1}{\sqrt{|k|^2 + s}} \nn
U_k \equiv - \frac{1}{\tau_2^2L^4} \frac{192 \pi^4 \epsilon^2}{(N+8)^2} \frac{1}{\left( |k|^2 + s \right)^{3/2}}
\end{gather}
For $N>1$, this is an anisotropic anharmonic oscillator.

\subsubsection{Two-particle Fock states}
\label{twoeffhams}

We now consider mixing between states of the form
\beq
\Psi_{\alpha\beta}[\varphi]|k_1,\alpha; k_2,\beta \rangle
\eeq
We have $|k_1,\alpha; k_2,\beta \rangle = S_{12}b^{\dagger}_{\alpha}(k_1) b^{\dagger}_{\beta}(k_2) |0 \rangle$ where we require a Bose symmetry factor: $S_{12} = \frac{1}{\sqrt{2}}$ if $k_1 = k_2$ and $\alpha = \beta$, and $S_{12} = 1$ otherwise. We will assume that there are no one-particle or $\geq 3$-particle states with overlap with this state. However, we will consider the case where there exists inequivalent states $|k_1,\alpha; k_2,\beta \rangle$, $|k_3,\alpha; k_4,\beta \rangle$, such that
\beq
\sqrt{|k_1|^2 + s} + \sqrt{|k_2|^2 + s} = \sqrt{|k_3|^2 + s} + \sqrt{|k_4|^2 + s}
\eeq
Because the interaction $V$ conserves momentum, these states will only mix if
\beq
k_1 + k_2 = k_3 + k_4.
\eeq
Such states can contribute to the low-energy spectrum on the torus. For example, on the square ($\tau = i$) torus the states $|2 \pi/L,\alpha; -2 \pi/L,\beta \rangle$ and $|2 \pi i/L,\alpha; -2 \pi i/L,\beta \rangle$ are inequivalent but can mix.

\begin{figure}
\centering
\includegraphics[width=4cm]{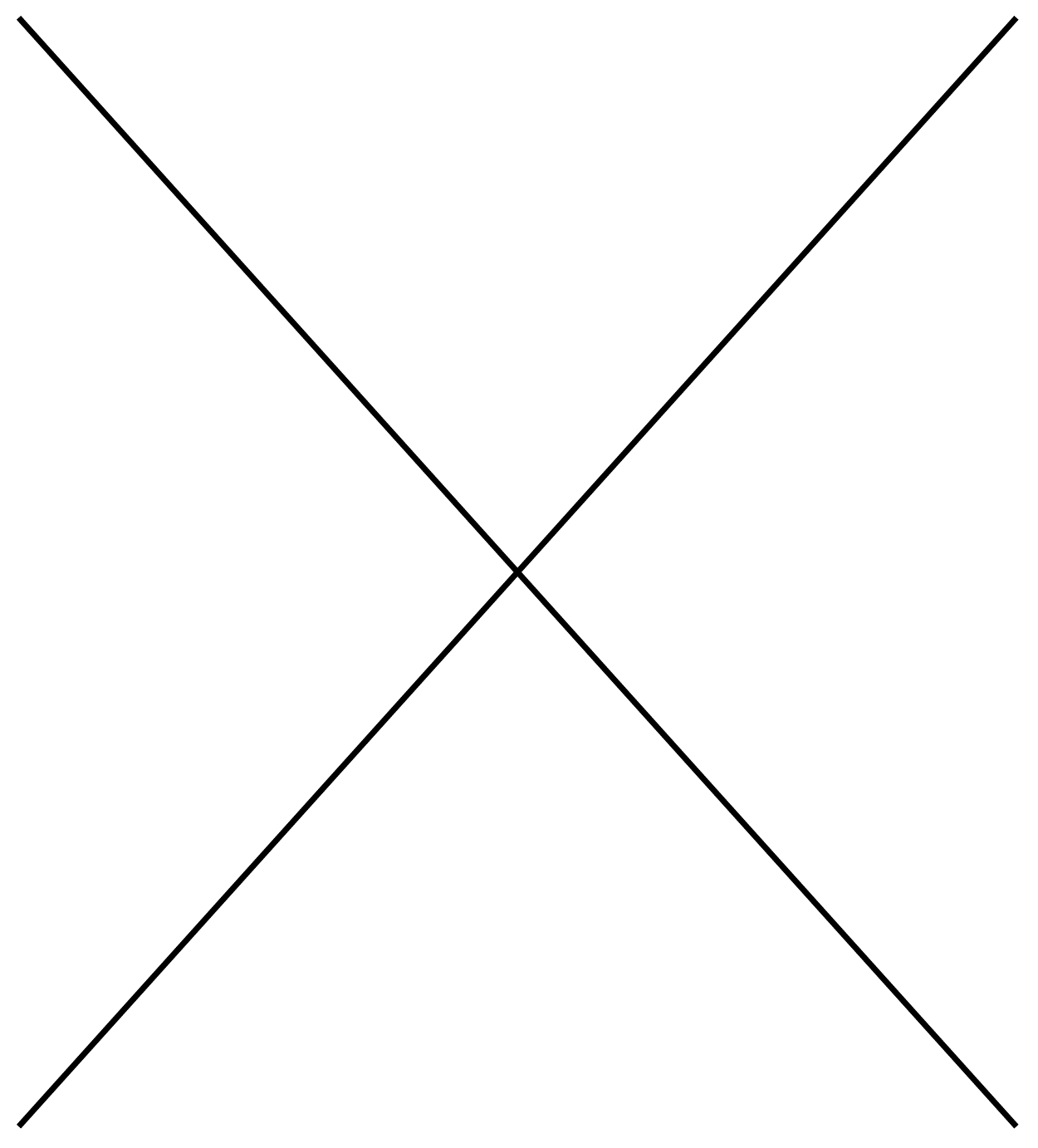}
\caption{The additional diagram which contributes to the splitting of the two-particle Fock states.}
\label{fig:twodiags}
\end{figure}

The effective Hamiltonian is now calculated in a similar manner to the previous two cases, and there is an obvious diagrammatic generalization of the previous rules. We now draw diagrams with four external lines with momenta $k_i$, $i = 1,...,4$. We then consider all possible contractions with either zero, one, or two vertices. We once again find a piece proportional to $h_{k=0}$, which involves the diagrams in Fig.~\ref{fig:zerodiags} but with the four external lines contracted and disconnected to the vertices. In addition, we get contributions which are simply the diagrams in Fig.~\ref{fig:singdiags} but with a single additional finite-momentum line disconnected from the rest. Finally, we obtain the additional diagram pictured in Fig.~\ref{fig:twodiags}, which can connect the inequivalent states considered above. This does not contain a loop, so it is finite.

An explicit calculation gives
\bea
h_{2k,\alpha \beta;\mu \nu} &=& \left[ \frac{\delta_{\alpha \mu}\delta_{\beta \nu}\delta_{k_1 k_3}\delta_{k_2 k_4} + \delta_{\alpha \nu} \delta_{\beta \mu} \delta_{k_1 k_4} \delta_{k_2 k_3}}{1 + \delta_{\mu \nu} \delta_{k_3 k_4}} \right] h_{k=0} \nn
&+& \frac{\delta_{\alpha \mu}\delta_{k_1 k_3}\delta_{k_2 k_4} }{1 + \delta_{\mu \nu} \delta_{k_3 k_4}} \left[ h_{k_3,\beta \nu} - \delta_{\beta \nu} h_{k=0} \right] + \frac{\delta_{\alpha \nu}\delta_{k_1 k_4}\delta_{k_2 k_3} }{1 + \delta_{\mu \nu} \delta_{k_3 k_4}} \left[ h_{k_4,\beta \mu} - \delta_{\beta \mu} h_{k=0} \right] \nn
&+& \frac{1}{\tau_2^{3/2} L^3} \frac{4 \pi^2 \epsilon}{(N+8)} \frac{\left(\delta_{\alpha \beta}\delta_{\mu \nu} + \delta_{\alpha \mu} \delta_{\beta \nu} + \delta_{\alpha \nu} \delta_{\beta \mu} \right)}{\left[ (|k_1|^2 + s)(|k_2|^2 + s)(|k_3|^2 + s)(|k_4|^2 + s) \right]^{1/2}}\frac{\delta_{k_1+k_2,k_3+k_4}}{(1 + \delta_{k_1 k_2} \delta_{k_3 k_4} \delta_{\alpha \beta}\delta_{\mu \nu})} \qquad
\label{twopartham}
\eea
The first two lines of Eq.~(\ref{twopartham}) can be given in terms of the the zero-particle and single-particle Hamiltonians, while the last term is new and contributes a constant shift in the energy. It is the last term which can mix two states unrelated by O($N$) symmetry, leading to a multidimensional Hamiltonian even when $N=1$. 

\subsubsection{Mixing between one- and two-particle Fock states}
\label{onetwomix}

We now consider the effective Hamiltonian which couples the states $|k,\alpha \rangle$ and $|k_1,\gamma;k_2,\delta \rangle$. In order for these states to mix, we need
\bea
&\sqrt{|k_1|^2 + s} + \sqrt{|k_2|^2 + s} = \sqrt{|k|^2 + s}&, \nn
&k_1 + k_2 = k_3.&
\eea
While this set of equations is very restrictive, when $s = 0$ it can be satisfied by choosing $k_1$ and $k_2$ to be collinear, so in our expressions where $\mathcal{A} s \lesssim \epsilon$, we always have $s \ll |k|^2$ and these states will mix. We will assume here that the only other mixing is due to O($N$) symmetry.

\begin{figure}
\centering
\includegraphics[width=4cm]{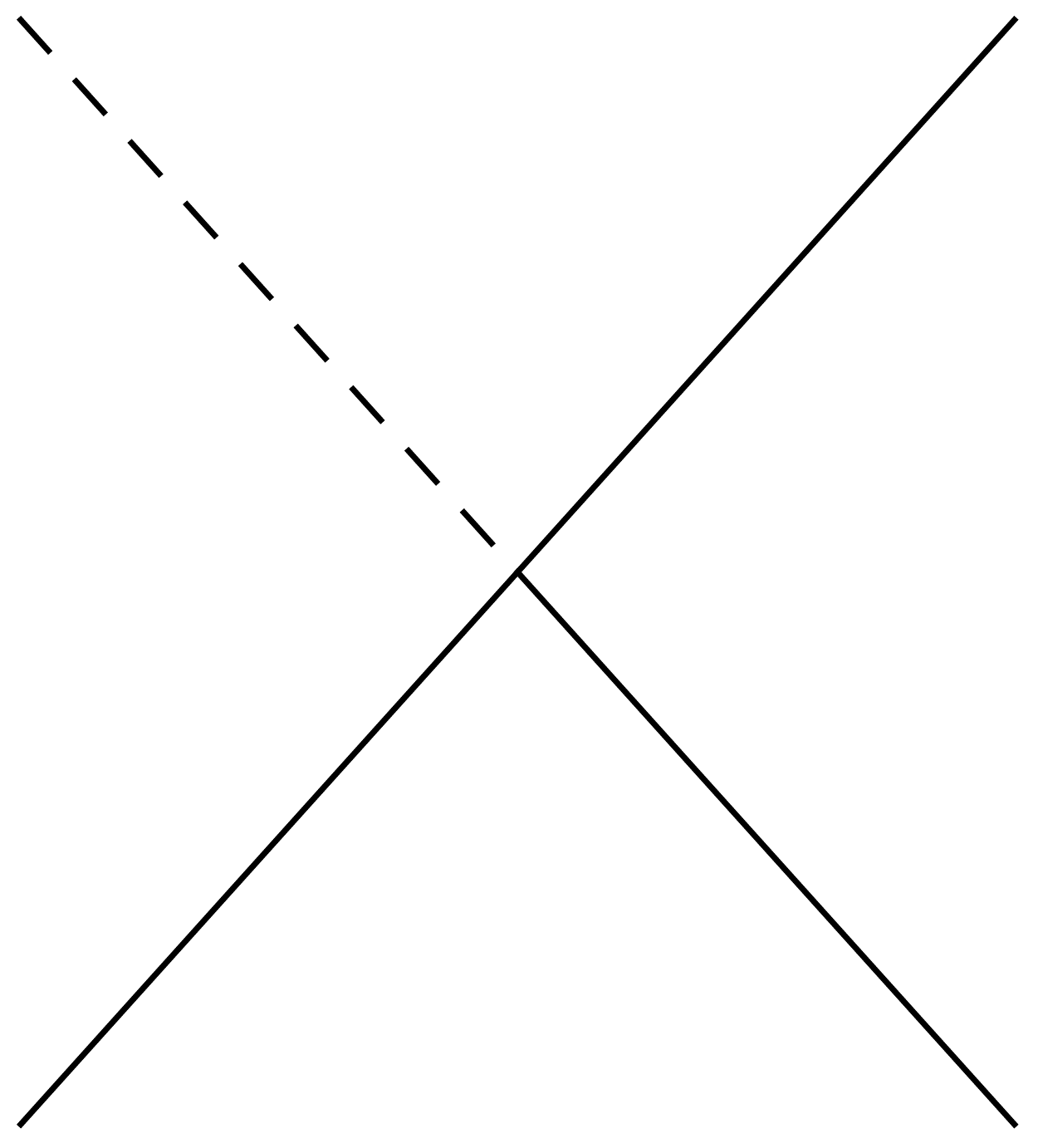}
\caption{The diagram which contributes to the effective Hamiltonian mixing single-particle and two-particle Fock states.
%\textcolor{red}{The dotted line is hard to distinguish from the other lines!}
}
\label{fig:mixdiags}
\end{figure}

We can write the Hamiltonian mixing these two states as
\beq
\begin{pmatrix}
h_{k,\alpha \beta} & h_{\alpha;\gamma \delta} \\
\left( h_{\beta;\mu \nu}\right)^{\dagger} & h_{2k,\mu \nu;\gamma \delta}
\end{pmatrix}
\begin{pmatrix}
|k,\beta \rangle \\
| k_1,\gamma;k_2,\delta \rangle
\end{pmatrix}
=
E_n \begin{pmatrix}
|k,\alpha \rangle \\
| k_1,\mu;k_2,\nu \rangle
\end{pmatrix}
\eeq
where the diagonal Hamiltonians are given by the one-particle and two-particle cases above, and the off-diagonal elements are
\beq
h_{\alpha;\gamma \delta} = \langle k, \alpha|V|k_1,\gamma, k_2,\delta \rangle - \langle k, \alpha|V \left( \frac{1 - P_{k}}{H_0 - \sqrt{|k|^2 + s} - \mathcal{E}_0} \right) V|k_1,\gamma, k_2,\delta \rangle
\eeq
where $P_k$ is the projector onto the degenerate subspace. We want all diagrams with three external lines with momenta $k$, $k_1$, and $k_3$. This leads to the diagram in Figure \ref{fig:mixdiags}, which is finite, giving
\beq
h_{\alpha;\gamma \delta} = \frac{1}{\sqrt{\tau_2} L} \frac{4 \sqrt{2} \pi^2 \epsilon}{(N+8)} \frac{\left( \delta_{\alpha \gamma}\varphi_{\delta} +  \delta_{\alpha \delta}\varphi_{\gamma} +  \delta_{\delta\gamma}\varphi_{\alpha} \right)}{\left( (|k|^2 + s)(|k_1|^2 + s)(|k_2|^2 + s) \right)^{1/4}}
\eeq
This coupling between Fock states with different particle number is another manifestation of the breakdown of well-defined quasiparticles at a quantum critical point.

At this point we hope it is clear how to generalize these results to more complicated cases. As one moves higher in the spectrum, more degeneracies between Fock states are possible and more complicated effective Hamiltonians will be needed.

\section{Numerical solution of the effective Hamiltonians}
\label{sec:num}

It remains to find the spectrum of the effective Hamiltonians perturbatively in $\epsilon$. This must be done numerically, since the $\epsilon$-expansion is equivalent to a strong-coupling expansion of a quartic anharmonic oscillator, which has no analytic solution. Given that there are many effective Hamiltonians, and each needs to be solved to at least third order in perturbation theory to obtain the $\epsilon^{4/3}$ contribution, this becomes one of the biggest barriers to obtaining the spectrum. 

Here we show how to obtain the spectrum for most of the low-lying effective Hamiltonians in the $N=1$ case, where the Hamiltonians are often one-dimensional quartic oscillators. For $N>1$, we will confine ourselves to solving the effective Hamiltonian which splits the Fock vacuum, where rotational invariance allows us to map it to a one-dimensional differential equation in ``radial'' coordinates. 

In Appendix \ref{app:largen} we discuss of the $N = \infty$ limit, where we can solve the spectrum of the Fock vacuum effective Hamiltonian exactly. We show agreement with the results of the large-$N$ expansion in Ref.~\cite{WS16} wherever possible. 
%In addition to these results, the $1/N$ corrections to the $N=\infty$ Hamiltonian can be obtained directly from known results on isotropic Hamiltonians \cite{AC90}.

\subsection{$N = 1$}
The results of this section were used in a previous publication by us \cite{HLSSW16}. We exploit the fact that most of the effective Hamiltonians for the low-lying states can be written in the form (see e.g. Eq.~(\ref{reducham}))
\beq
H_{\mathrm{eff}} = \mathcal{E} + u h[r]
\eeq
where
\beq
h[r] = -\frac{1}{2} \frac{d^2}{d\varphi^2} + \frac{r}{2} \varphi^2 + \varphi^4 
\eeq
Here, $\mathcal{E}$ is a known function of $\epsilon$, and
\bea
u &=& u_1 \epsilon^{1/3} + u_2 \epsilon^{4/3} + \mathcal{O}(\epsilon^{7/3})\nn
r &=& r_1 \epsilon^{1/3} + \mathcal{O}(\epsilon^{4/3})
\eea
for some constants $r_{1}$, $u_{1,2}$ which depend on the specific effective Hamiltonian in question. Then we need the spectrum of $h[r]$ in a power series in $r$. After writing
\beq
h[r] \Psi[\varphi] = \lambda_n(r) \Psi[\varphi],
\eeq
expansion of the spectrum in $\epsilon$ can be written
\beq
\lambda_n(r) = \sum_{m = 0}^{\infty} c_{n,m} r^m 
\label{strongcup1}
\eeq
Assuming we know the coefficients $c_{n,m}$ which appear in this expansion, we can simply write down the spectrum to the desired order:
\beq
E_n(\epsilon) = \mathcal{E} + u_1 c_{n,0} \epsilon^{1/3} + u_1 r_1 c_{n,1} \epsilon^{2/3} + u_1 r_1^2 c_{n,2} \epsilon + \left( u_1 r_1^3 c_{n,3} + u_2 c_{n,0} \right) \epsilon^{4/3} + \mathcal{O}\big(\epsilon^{5/3}\big)
\label{specising}
\eeq
Fortunately, the numerical calculation of the coefficients $c_{n,m}$ has been extensively studied in the literature. In Ref.~\cite{SCZ99}, Tables 7 and 10, the coefficients $c_{n,m}$ for $m = 1,..,10$, $n=1,...,10$ are given with at least five digit accuracy. Thus, for one-dimensional effective Hamiltonians, the spectrum can be obtained.

\begin{table}
\begin{tabular}{| c || c | c | c | c |}
\hline
$\tau = i$ & $\kappa = 0$ & $\kappa = 1$ & $\kappa = \sqrt{2}$ & $\kappa = 2$  \\
\hline
\multirow{10}{*}{$L E$} & -0.12 & & & \\
&\cellcolor[gray]{.85}1.70 & & & \\
& 5.04 & & & \\
& & 6.28\cellcolor[gray]{.85} & & \\
& &8.76 & & \\
& &  &\cellcolor[gray]{.85}8.90 & \\
&\cellcolor[gray]{.85}8.90 & & & \\
& & & 11.23 & \\
& & \cellcolor[gray]{.85}12.29& & \\
& & & & \cellcolor[gray]{.85}12.55 \\
\hline
\end{tabular}
\caption{Low-lying spectrum of the critical Ising model on the square torus from $\epsilon$-expansion. The states shaded gray are odd under the global $\mathbb{Z}_2$ symmetry, while the unshaded states are $\mathbb{Z}_2$ even. Here we parametrize the momentum by $\kappa = L|k|/(2 \pi)$.}
\label{tab:critising}
\end{table}

However, there can still be multi-dimensional effective Hamiltonians for $N=1$ due to mixing between different Fock states. Fortunately, the form of the mixing is often very simple. In computing the low-lying spectrum, the first mixing one comes across is between two inequivalent two-particle states, as described in Section \ref{twoeffhams}. There, the effective Hamiltonian acting on these two states takes the form
\beq
h_{2k} = h_{2k,1}\mathbb{I} + h_{2k,2} \sigma^x
\eeq
where $h_{2k,2}$ is of order $\epsilon$ and is independent of the zero mode. In this case, we can diagonalize the Hamiltonian by inspection, obtaining the energy splitting
\bea
E_{2k,-} &=& E_{2k,1} - h_{2k,2} \nn
E_{2k,+} &=& E_{2k,1} + h_{2k,2}
\eea
where $E_{2k,1}$ are the energies of $h_{2k,1}$.

Finally, we do encounter the states described in Section \ref{onetwomix} which mix the single-particle and two-particle Fock states. In this case, we must numerically calculate the contribution. However, this is made easier by the fact that the order $\epsilon^{5/6}$ term arises in first-order perturbation theory, so we only need the zeroth order wave functions. That is, we first numerically calculate the zeroth order wave functions which diagonalize $h_k$ and $h_{2k}$, and then we compute the overlap
\beq
\int_{-\infty}^{\infty} d \varphi \Psi_{k}[\varphi] \Psi_{2k}[\varphi] h_{1k2}
\eeq
We calculate the unperturbed wave function numerically using the same shooting method described in Appendix \ref{numerics} for the $N>1$ case.

The low-lying spectrum of the critical Ising model is given in Table \ref{tab:critising}.

\subsection{$N>1$}

Here we will focus on the splitting of the Fock vacuum, where the Hamiltonian is an isotropic $N$-dimensional oscillator. We begin with Eq.~(\ref{reducham}):
\beq
h_{k=0} = \mathcal{E}_{k = 0} + \frac{1}{\sqrt{\tau_2} L} \left( \frac{U}{4!} \right)^{1/3} \left( \frac{\pi_{\alpha}^2}{2} + \frac{RU^{-2/3}}{2} \varphi_{\alpha}^2 + \varphi_{\alpha}^4 \right)
\eeq
Then defining
\beq
h_N[s] = -\frac{1}{2} \nabla_{\varphi}^2 + \frac{s}{2} \varphi_{\alpha}^2 + \varphi_{\alpha}^4
\label{radialeq}
\eeq
and the reduced couplings
\bea
\frac{1}{\sqrt{\tau_2} L} \left( \frac{U}{4!} \right)^{1/3} &=& u_1 \epsilon^{1/3} + u_2 \epsilon^{4/3} + \mathcal{O}(\epsilon^{7/3}) \nn
RU^{-2/3} &=& r_1 \epsilon^{1/3} + \mathcal{O}(\epsilon^{4/3}),
\eea
we obtain Eq.~(\ref{specising}) except the coefficients $c_{n,m}$ will depend on $N$. 

To make further progress, we take advantage of the fact that $h_N[r]$ can be written in spherical coordinates, after which it separates into a known angular equation and a one-dimensional radial equation. We go to hyperspherical coordinates
\bea
\varphi_1 &=& \rho \cos \nn
\varphi_1 &=& \rho \sin \theta_1 \cos \theta_2  \nn
\varphi_1 &=& \rho \sin \theta_1 \sin \theta_2 \cos \theta_3 \nn
&&\vdots \nn
\varphi_{N-1} &=& \rho \sin \theta_1 \cdots \sin \theta_{N-2} \cos \theta_{N-1} \nn
\varphi_N &=& \rho \sin \theta_1 \cdots \sin \theta_{N-2} \sin \theta_{N-1}
\label{hypersph}
\eea
Then the Laplacian can be written in the separable form
\beq
\nabla_{\varphi}^2 = \sum_{i = 1}^N \frac{\partial^2}{\partial^2 \varphi_i} = \frac{1}{\rho^{N-1}} \frac{\partial}{\partial \rho} \rho^{N-1} \frac{\partial}{\partial \rho} + \frac{1}{\rho^{N-1}} \nabla_{S^{N-1}}^2
\eeq
where $\nabla_{S^{N-1}}^2$ is the Laplacian on the $(N-1)$-sphere. We will not use the coordinate representation of this operator, but instead use what is known of its spectrum \cite{SW71}. The eigenvectors and eigenvalues are given by
\beq
-\nabla_{S^{N-1}}^2 Y_{\ell,\ell_1,\ell_2,...,\ell_{N-2}}(\theta_i) = \ell \left( \ell + N - 2 \right) Y_{\ell,\ell_1,\ell_2,...,\ell_{N-2}}(\theta_i)
\eeq
where the eigenfunctions can be given in terms of Gegenbauer polynomials of the $\cos \theta_i$, and the indices can range from
\bea
\ell &=& 0, 1, 2, ... \nn
\ell_1 &=& -\ell_2, -\ell_2 + 1,-\ell_2 + 2,..., \ell_2 \nn
\ell_2 &=& 0, 1, 2, ..., \ell_3 \nn
\ell_3 &=& 0, 1, 2, ..., \ell_4 \nn
&&\vdots \nn
\ell_{N-2} &=& 0, 1, 2,...,\ell
\eea
The spectrum does not depend on the the $\ell_i$. This gives a degeneracy for a given eigenvalue $\ell$ of
\beq
\mathcal{N}(\ell,N) = \frac{(2 \ell + N - 2)(\ell + N - 3)!}{\ell! (N-2)!}
\eeq
for $N \geq 3$. For $N=2$, there is only one eigenfunction for each $\ell$, but the states $Y_{\ell}$ and $Y_{-\ell}$ are degenerate (these are simply the states $e^{i \ell \theta}$ and $e^{-i \ell \theta}$), so the degeneracy is
\beq
\mathcal{N}(\ell,2) = \begin{cases} 1 \qquad \ell = 0\\ 2 \qquad \ell > 0 \end{cases}
\eeq
Finally, we note that the eigenfunctions are in the symmetric traceless tensor representation of O($N$), and we can label these representations using the eigenvalue $\ell$.

With these eigenfunctions, we can express our functionals as
\beq
\Psi[\varphi] = R_{n,\ell}(\rho) Y_{\ell,\ell_1,\ell_2,...,\ell_{N-2}}(\theta_i)
\eeq
and the eigenvalue equation becomes
\beq
\frac{1}{2} \left( - \frac{1}{\rho^{N-1}} \frac{\partial}{\partial r} \rho^{N-1} \frac{\partial}{\partial \rho} - \frac{\ell \left( \ell + N - 2 \right)}{\rho^{N-1}} + r \rho^2 + 2 \rho^4 \right) R_{n,\ell}(\rho) = E_{n,\ell} R_{n,\ell}(\rho)
\eeq
Here we have introduced a radial quantum number $n$, which corresponds to the number of zeros in $R$. We have reduced our problem to a one-dimensional eigenvalue equation, and we wish to find the spectrum perturbatively in $r$. In analogy with the $N=1$ case, we write this expansion as
\beq
E_{n,\ell} = \sum_{m = 0}^{\infty} c_{n,\ell,m} r^m
\eeq

\begin{table}
\begin{tabular}{| c || c | c | c | c | c |}
\hline
$\tau = i$ & $\ell = 0$ & $\ell = 1$ & $\ell = 2$ & $\ell = 3$ & $\ell = 4$ \\
\hline
\multirow{10}{*}{$L E$} & -0.986 & & & & \\
& & 0.96& & & \\
& & & 3.66 & & \\
& 4.74 & & & & \\
& & & & 6.79 & \\
& & 8.31 & & & \\
& & & & &10.32 \\
& & & 12.12 & & \\
& & & & 16.19 & \\
& & & & & 20.52 \\
\hline
\end{tabular}
\caption{Low-lying spectrum of the critical O(2) model on the square torus from $\varepsilon$-expansion, including the ground state energy. These states are obtained from the effective Hamiltonian which gives the splitting of the $k=0$ Fock vacuum. The $\ell = 0$ states are non-degenerate while the $\ell > 0$ states are two-fold degenerate.}
\label{tab:2critsq}
\end{table}

\begin{table}
\begin{tabular}{| c || c | c | c | c | c |}
\hline
$\tau = i$ & $\ell = 0$ & $\ell = 1$ & $\ell = 2$ & $\ell = 3$ & $\ell = 4$ \\
\hline
\multirow{10}{*}{$L E$} & -2.28 & & & & \\
& & -0.66& & & \\
& & & 1.74 & & \\
& 2.95 & & & & \\
& & & & 4.59 & \\
& & 6.20 & & & \\
& & & & &7.81 \\
& & & 9.69 & & \\
& & & & 13.45 & \\
& & & & & 17.47 \\
\hline
\end{tabular}
\caption{Low-lying spectrum of the critical O(3) model on the square torus from $\varepsilon$-expansion, including the ground state energy. These states are obtained from the effective Hamiltonian which gives the splitting of the $k=0$ Fock vacuum. The states have degeneracy $2 \ell + 1$.}
\label{tab:3critsq}
\end{table}

We have obtained the coefficients of the perturbative expansion in $r$ for $N=2,3,4$, $\ell = 0,...,4$ and $n = 0,1$ numerically. We obtained these by first solving the $r=0$ equation numerically for the wave function and energy. We then used logarithmic perturbation theory \cite{AA79a,AA79b}, which is well-suited to this problem because it allows one to find the coefficients of the expansion directly from the unperturbed energy and wave function without needing excited states. We discuss our approach in Appendix \ref{numerics}, and give the coefficients of the expansion in Tables \ref{tab:n=2}-\ref{tab:n=4}. Using these results, the energy to leading order is
\beq
E_{n,\ell}(\epsilon) = \mathcal{E} + u_1 c_{n,\ell,0} \epsilon^{1/3} + u_1 r_1 c_{n,\ell,1} \epsilon^{2/3} + u_1 r_1^2 c_{n,\ell,2} \epsilon + \left( u_1 r_1^3 c_{n,\ell,3} + u_2 c_{n,\ell,0} \right) \epsilon^{4/3} + \mathcal{O}\big(\epsilon^{5/3}\big)
\label{specon}
\eeq
We give the lowest-lying states for $N=2$ and $N=3$ at the critical point in Tables \ref{tab:2critsq} and \ref{tab:3critsq}.

\section{Numerical calculation of critical torus spectra from lattice models}
\label{sec:ED}

In this section we investigate the critical torus spectra numerically using
Exact Diagonalization (ED). We study explicit lattice models known to exhibit a
transition in the O($N$) universality class, compute their spectrum on finite
clusters and extrapolate it to the thermodynamic limit. 
We then compare these numerical results to the analytic calculations from
the $\epsilon$-expansion. 
In the following, we will present results for $N=2$ and $N=3$ and refer to
Ref.~\cite{HLSSW16} for a detailed discussion about the $N=1$ (Ising) CFT.

\subsection{$N = 2$}
We first consider the O(2) critical point, also known as the quantum XY model.
Two different lattice models will be utilized to study the critical spectrum of
this universality class numerically. The first is a spin-1 model, $S=1$, with single-ion
anisotropy~\cite{Zhang2013,Stoudenmire2014} 
\begin{equation}
H_1^{\text{O(2)}} = -J \sum_{\langle i,j \rangle} \left( S_i^x S_j^x + S_i^y S_j^y \right)
- J_{z} \sum_{\langle i,j \rangle} S_i^z S_j^z
+ D \sum_i (S_i^z)^2
\label{eq:O2_spin1}
\end{equation}
We set the energy scale by choosing $J=1$.  For small $D$ the system orders
ferromagnetically in the $x$-$y$ spin plane, while for large $D$ the system
approaches a product state of single spins with $S^z=0$.  The phase transition
is found to be in the XY/O(2) universality class. The parameter $J_z$ can be
tuned within a range around zero to check the stability of our results.

The second model we consider is the spin-1/2 XY-bilayer
model~\cite{Stoudenmire2014,Helmes2014}. It consists of two usual
ferromagnetic XY layers with additional XY couplings between them. We denote a
spin located on site $i$ in the first (second) layer as $S_{i,1(2)}$, the model
is then described by
\begin{equation}
 H_2^{\text{O(2)}} = -J \sum_{n=1}^2 \sum_{\langle i,j \rangle} \left( S_{i,n}^x S_{j,n}^x +
  S_{i,n}^y  S_{j,n}^y \right)
  + J_{\perp} \sum_i \left(S_{i,1}^x S_{i,2}^x + 
  S_{i,1}^y S_{i,2}^y \right)
\label{eq:O2_bil}
\end{equation}
We only consider positive couplings $J,J_{\perp}>0$ here and set our energy scale
$J=1$. For large $J_{\perp}$ the system is described by a product state of singlets on
each interlayer bond, whereas a XY-ferromagnet is formed in each plane for small
$J_{\perp}$. The two phases are separated by a XY quantum critical point at $J_{\perp} =
J_{\perp}^c = 5.460(1) J$~\cite{Helmes2014}\footnote{The energy spectrum of this model
    is conserved under changing the sign of $J \rightarrow -J$ when the momenta
    for odd $S^z$ levels are shifted as $\mathbf{k} \rightarrow \mathbf{k} +
    (\pi,\pi)$. Therefore, the critical coupling $J_{\perp}^c$ is identical for both,
    ferromagnetic and antiferromagnetic $J=\pm1$.}

To calculate the critical torus energy spectrum, we compute the spectrum of the
Hamiltonians at criticality on finite-size toric clusters numerically.  We
multiply the finite-size spectra with the linear system size $L$ to get rid of
the dominant scaling and then extrapolate these spectra in $1/N$ to the
thermodynamic limit. Further details about the numerical approach can be found
in App.~\ref{app:ED}.

\begin{figure}[ht]
 \centering
 \includegraphics[height=7cm]{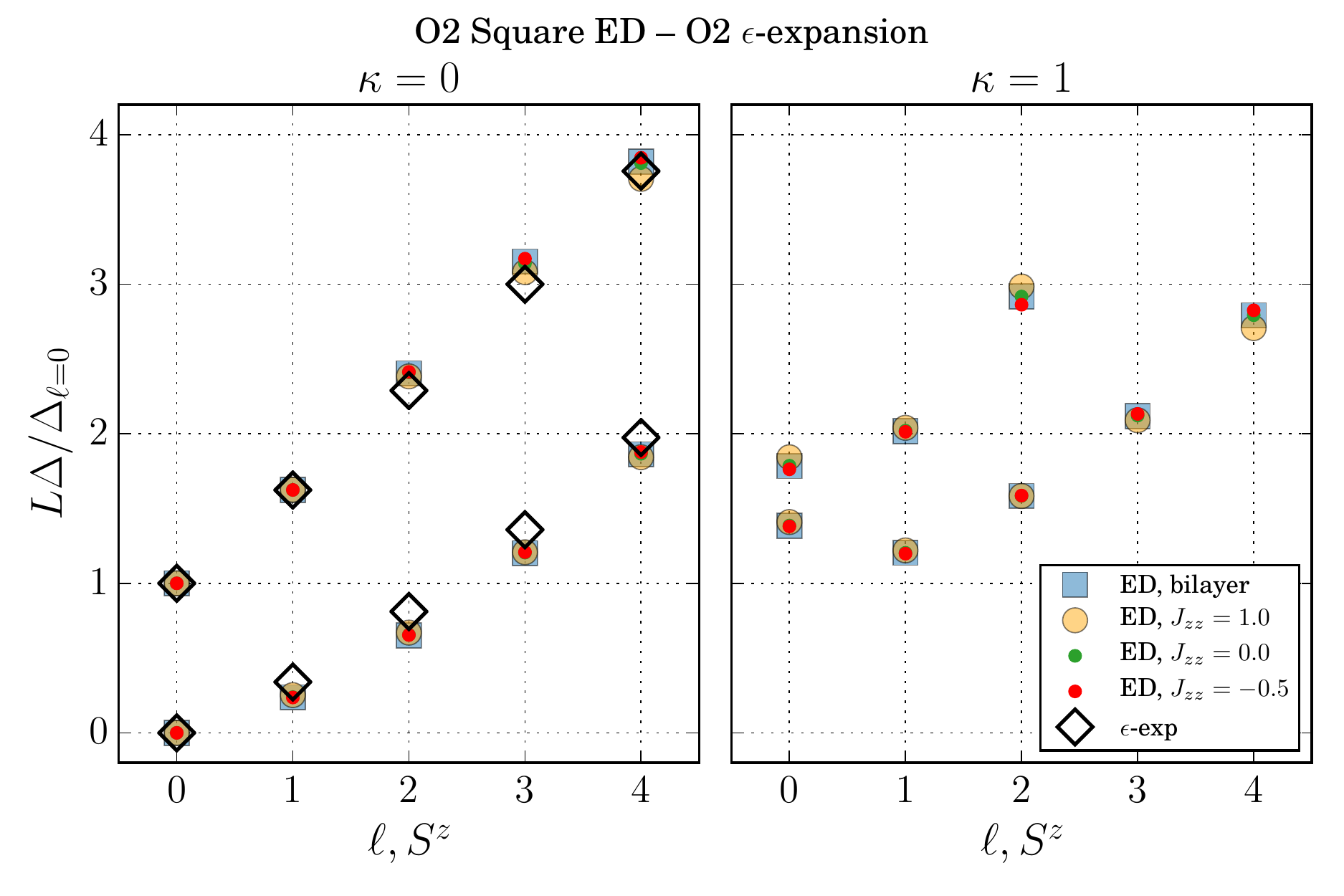}
 \caption{Critical torus energy spectrum for the O(2) CFT on a square geometry
     $\tau=i$ in the $\kappa=0$ (left) and the $\kappa=1$ (right) sectors.  The
     results are normalized such that $\Delta_{\ell=0}=1$. Full symbols denote
     ED results for different models/parameters, empty black diamonds show the
     $\epsilon$-expansion results for the critical O(2) CFT ($\kappa=0$ only).
     $\ell=0$ levels are non-degenerate, while $\ell>0$ levels are two-fold
     degenerate (times the geometrical multiplicity of $\kappa>0$ levels). Note
     that the ED results only show levels in the fully symmetric representation
     regarding the lattice point-group symmetry. Levels in other point-group
     representations start to appear above $\Delta/\Delta_{l=0} \gtrsim 2$ for
     $\kappa=0$.}
 \label{fig:O2_critical}
\end{figure}

In Fig.~\ref{fig:O2_critical} we present the numerically obtained critical O(2)
torus energy spectrum in the $\kappa=0$ and $\kappa=1$ sectors.  Here and in the
following we only show the lowest energy levels for $\ell\leq4$ which are in the
fully symmetric representation regarding the lattice point-group symmetry. We
also restrict our discussion to square lattices, $\tau=i$. Results for
triangular geometry $\tau=\frac{1}{2}+\frac{\sqrt3}{2}i$ are given in
App.~\ref{app:ED}.
The spectrum is normalized such that the lowest gap in the $\ell=0$ sector is
set to $\Delta_{\varepsilon_T} \equiv \Delta_{\ell=0} = 1$. To demonstrate the
stability of the numerical results and the universality of the spectrum, the
different models and parameters considered for O(2) universality are shown with
different symbols and colors in the plot.
For $\kappa=0$ we also plot the $\epsilon$-expansion results with empty diamonds to
compare them with the numerics. We find that they agree reasonably well and show
a qualitatively identical structure.
This highlights, that the critical torus energy spectrum is a universal
fingerprint of the underlying CFT which is available from a wide variety of
analytical and numerical approaches.

Here we also want to note, that the four relevant fields in the O(2)
CFT~\cite{Pelissetto2002} correspond to the lowest $\ell=1$, $\ell=2$ and
$\ell=3$ levels as well as the first excited $\ell=0$ level in the critical
torus spectrum (all $\kappa=0$). Interestingly, these are the four lowest
states in the spectrum. Also in the Ising CFT, the two relevant fields
correspond to the two lowest levels in the critical spectrum~\cite{HLSSW16}.
This indicates, that it might be a general feature that relevant fields of the
CFT have light analogues in the critical torus spectrum.

The numerical values for the critical torus spectrum are tabulated in
App.~\ref{app:ED} in Tab.~\ref{tab:critO2EDsquare} for the square geometry
$\tau=i$ and in Tab.~\ref{tab:critO2EDtriangular} for the triangular geometry
$\tau=\frac{1}{2} + \frac{\sqrt{3}}{2}i$.

\subsection{$N=3$}

In this section, we study the critical torus spectrum of the O(3) CFT
numerically. To do so, we again consider two different lattice models with a
critical point known to be described by the O(3) universality class.
The first model is the prototypical Heisenberg bilayer model
\cite{Sandvik1994,Wang2006,Lohofer2015,Nishiyama2016}. It consists of spin-1/2 on
two layers with nearest-neighbour Heisenberg intraplane couplings and Heisenberg
interplane couplings on the rungs:
\begin{equation}
    H_1^{O(3)} = J \sum_{n=1}^2 \sum_{\langle i,j \rangle} \left( \mathbf{S}_{i,n} \cdot
\mathbf{S}_{j,n} \right) + J_2 \sum_i \mathbf{S}_{i,1} \cdot \mathbf{S}_{i,2}
\label{eq:O3_bilayer}
\end{equation}
$S_{i,n}$ denotes a spin on site $i$ in layer $n$. 
We set ferromagnetic intraplane couplings $J=-1$ and
antiferromagnetic rung couplings $J_2>0$. For large $J_2$ the groundstate
is a product state of singlets on each rung, whereas the
groundstate for small $J_2$ is the direct product of Heisenberg ferromagnets within each plane. The
phases are separated by an O(3) critical point.

The second model we want to investigate here is a Heisenberg model on a 2D square
lattice with columnar dimerization of bonds \cite{Wenzel2009,Matsumoto2001}. The
Hamiltonian for this ladder model is
\begin{equation}
 H = J \sum_{\langle i,j \rangle} \mathbf{S}_i \cdot \mathbf{S}_j 
     +J_2 \sum_{\langle i,j \rangle^{\prime}} \mathbf{S}_i \cdot \mathbf{S}_j
\label{eq:O3_2D}
\end{equation}
Every second horizontal bond on the lattice is chosen to be in the family
$\langle i,j \rangle^{\prime}$, such that these dimerized bonds form ladders
and every spin is part of exactly one dimerized bond. We set all couplings
antiferromagnetic and set the energy scale $J=1$.  For $J_2/J=1$ a N\'eel AFM
is stabilized on a square lattice and for large $J_2/J$ a product state of
singlets on the bonds $\langle i,j \rangle^{\prime}$ is formed as a
groundstate. These phases are separated by an O(3) transition at the critical
coupling $(J_2/J)_c = 1.9096(2)$~\cite{Wenzel2009}.

\begin{figure}[h]
 \centering
 \includegraphics[height=7cm]{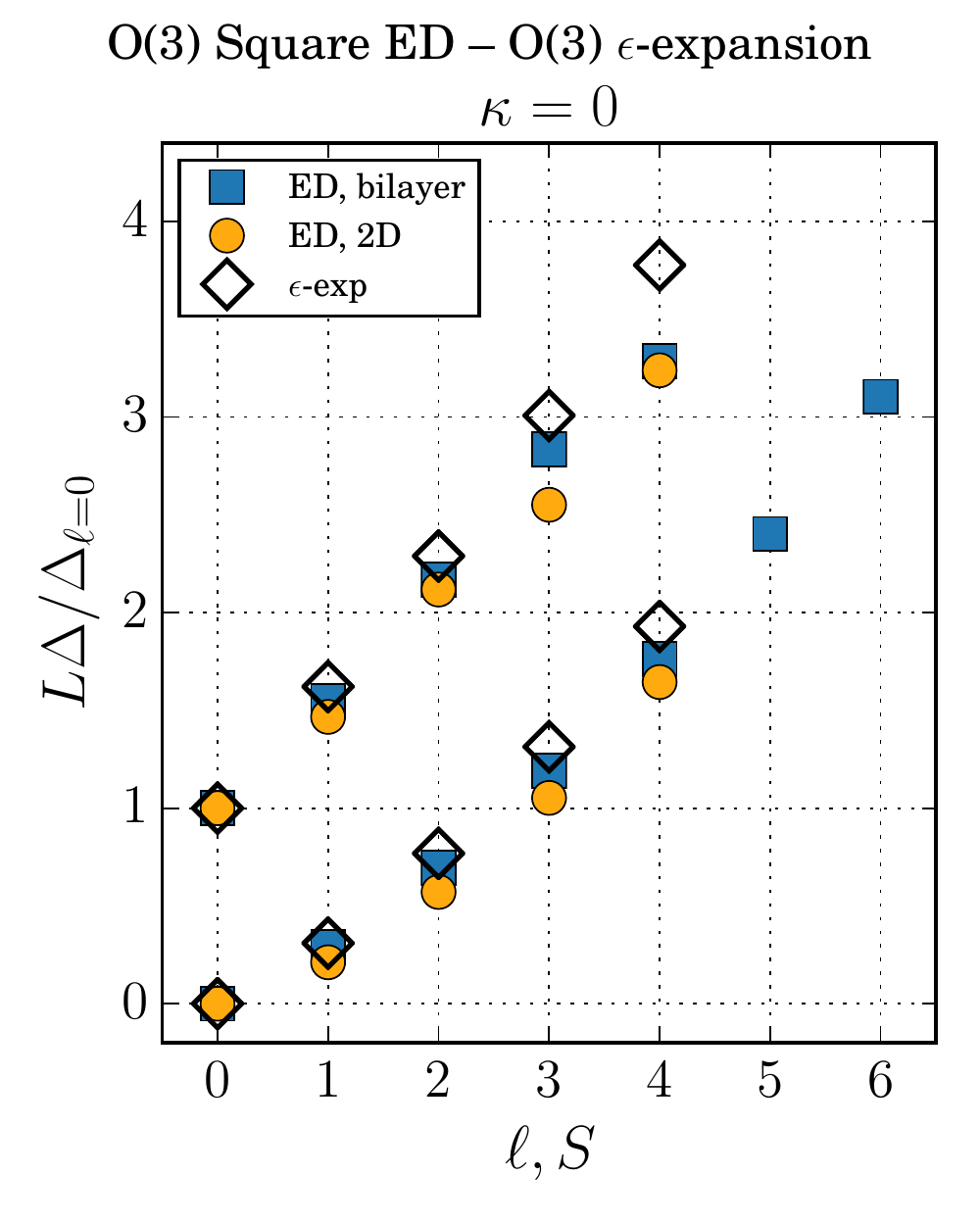}
 \caption{Critical torus energy spectrum for the O(3) CFT on a square lattice in
     the $\kappa=0$ sector obtained from the models \eeqref{eq:O3_bilayer} (blue
     squares) and \eeqref{eq:O3_2D} (yellow circles). The results are normalized
     such that $\Delta_{l=0}=1$.  Empty black diamonds show the
     $\epsilon$-expansion results for the O(3) CFT.  The levels are $2l+1$-fold
     degenerate. Note that the ED results only show levels in the fully
     symmetric representation regarding the lattice point-group symmetry.
     Levels in other point-group representations start to appear above
     $\Delta/\Delta_{l=0} \gtrsim 2$ for $\kappa=0$.}
 \label{fig:O3_critical}
\end{figure}

We proceed similar to the case of $N=2$ to compute the critical torus energy
spectrum for the O(3) CFT numerically. The critical spectrum is shown in
Fig.~\ref{fig:O3_critical} together with the $\epsilon$-expansion results. We
again observe a decent agreement between the two methods and qualitatively
identical critical torus spectra. Larger deviations for the second level in
$\ell=4$ are probably related to difficulties in the extrapolation to the
thermodynamic limit, as the available system sizes are strongly limited for
these models.
Although the critical torus spectra in the $\kappa=0$ sector seems to look very
similar for the O(2) and O(3) CFTs, their degeneracy structure is inherently
different. For O(3) the levels are $(2\ell+1)$-fold degenerate, whereas they are
2-fold (1-fold) degenerate for $\ell>0$ ($\ell=0$) in the O(2) CFT. The
degeneracy structure is thus an important feature, yielding qualitatively
very distinct critical torus spectra.

The numerical values for the critical torus spectrum of the O(3) CFT from ED
are tabulated in App.~\ref{app:ED} in Tab.~\ref{tab:critO3EDsquare} for the
square geometry $\tau=i$. In Tab.~\ref{tab:critO3EDtriangular} we additionally
list the energy levels for the triangular geometry $\tau=\frac{1}{2} +
\frac{\sqrt{3}}{2}i$.

\section{Conclusions}
\label{sec:conc}

\begin{figure}[h]
 \centering
 \includegraphics[height=7cm]{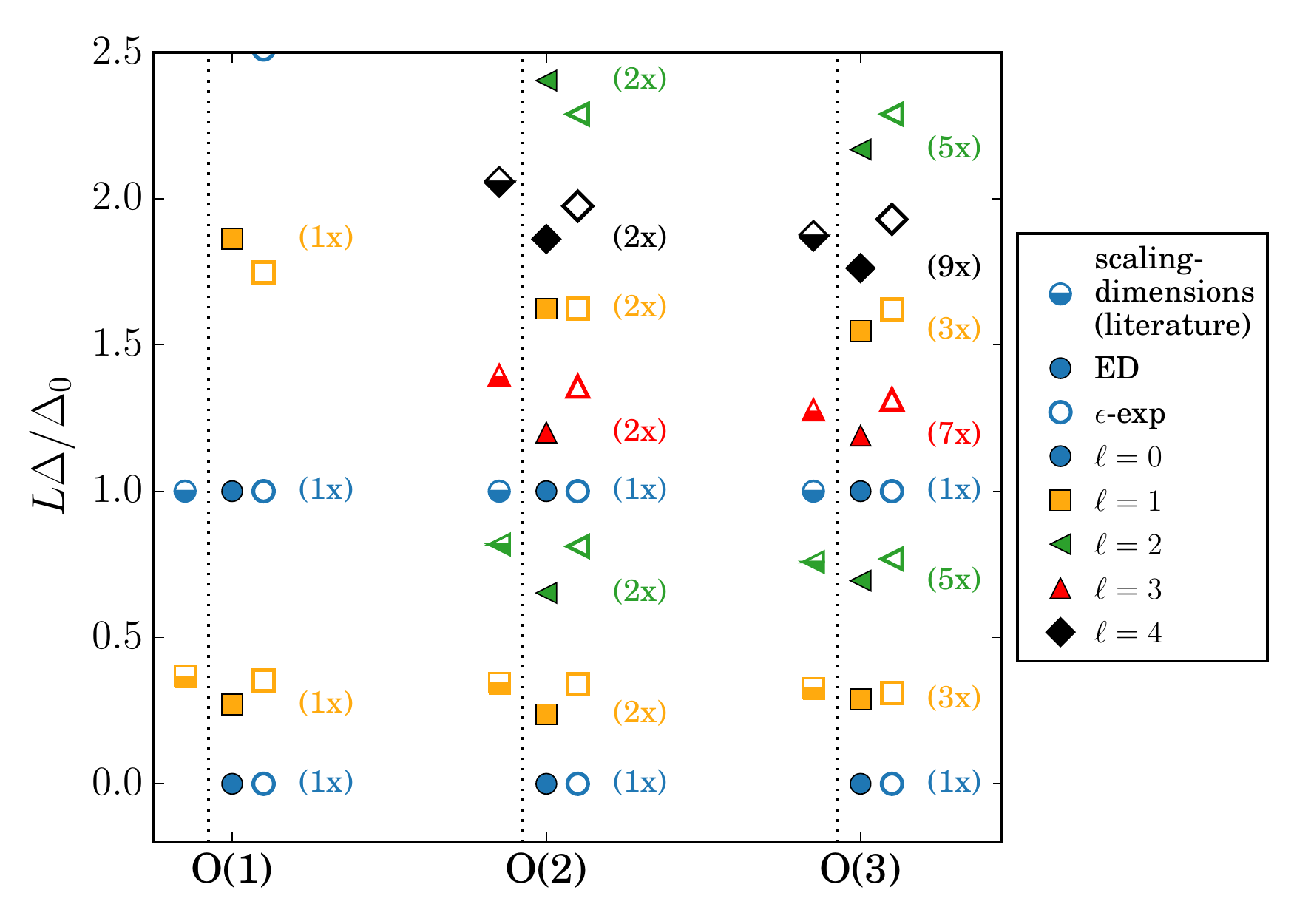}
 \caption{Critical low-energy torus spectra for the discussed O($N$) CFTs for
     $\kappa=0$ and $\ell \leq4$ compared to the operator scaling dimensions of
     the CFTs from Refs.~\cite{El-Showk2012,Hasenbusch2011}. 
     The spectra and scaling dimensions are normalized by the gap $\Delta_0$ to
     the first excited state with $\ell=0$. 
     Full symbols denote results from numerics (ED), open symbols show
     $\epsilon$-expansion results.  Half-filled symbols show the operator
     scaling dimensions of the corresponding CFTs as a comparison.
     The different symbols represent the different values of $\ell$. The numbers
     in parentheses give the degeneracy of the levels.
     The level structure including degeneracies is qualitatively different
     between the distinct CFTs and can be considered as a universal fingerprint
     of the CFT. The operator scaling dimensions correspond to the critical
     energy spectrum of the Hamiltonians on a sphere. Interestingly, the
     structure of the operator dimensions and the torus spectrum are very similar
     for the low levels with an additional low $\ell=1$ level in the torus spectrum.
 }
 \label{fig:ON_comparison}
\end{figure}

In this paper, we have demonstrated how to compute the torus energy spectrum of the Wilson-Fisher CFT from quantum field theory in the $\epsilon$-expansion, showing the expected emergence of a spectrum determined entirely by the universality class of the CFT and the spatial geometry of the torus. 
We have also computed the spectrum of several explicit lattice models at their
respective quantum critical points using ED, and have shown that the analytic
and numerical calculations agree well with each other, highlighting that the
finite-size spectrum is a useful universal fingerprint for identifying quantum
criticality. In Fig.~\ref{fig:ON_comparison} we show a comparison of the
critical torus low-energy spectra for $N=1,2,3$ in the $\kappa=0$ sector which
substantiates this interesting aspect of a universal fingerprint. Additionally,
we compare the critical torus spectra with operator scaling dimensions of the
corresponding CFTs from literature~\cite{El-Showk2012,Hasenbusch2011}. We note a quantitative match between the low-energy critical torus spectrum at $\kappa=0$  and the (rescaled) operator
dimensions. This match may be a coincidence, as there is no known mapping between the torus spectrum and the operator dimension. It is not known what constitutes the complete data for a three-dimensional CFT, and whether the torus spectrum is new data or whether is can be related to the set of operator scaling dimensions. 

We hope that our work has shed light on the nature of the finite-size spectrum in CFTs where conformal invariance does not lead to simple and exact results, and that our calculations aid in identifying critical behavior in numerical studies of quantum lattice models by investigation of the critical energy spectrum. The methods used in this work should also be applicable to computing the finite-size spectra of other CFTs, such as an $\epsilon$-expansion approach to the Gross-Neveu-Yukawa CFT.

Another application of the present results is to a class of deconfined quantum critical points known as the O($N)^\ast$ models. These are described by the same field theory as the O($N$) model except that the order parameter carries a $\mathbb{Z}_2$ gauge redundancy, imposing an equivalence relation $\phi \sim -\phi$ \cite{RJSS91,CSS93}. 
In these models, the order parameter $\phi$ can be thought of as the spinon in a $\mathbb{Z}_2$ spin liquid, whose condensation results in a confinement phase transition accompanied by broken symmetry associated with the internal O($N$) quantum number.

The prescription for relating our current results to the O($N)^\ast$ model was detailed in Refs.~\cite{HLSSW16,WS16}, and amounts to a selection rule on the allowed $\mathbb{Z}_2$ parity of the spectrum and on the inclusion of states with twisted boundary conditions $\phi(x + iy + n_1 \omega_1 + n_2 \omega_2) = \pm \phi(x + iy)$. 
The spectrum of states with twisted boundary conditions is easily computed with $\epsilon$-expansion due to the absence of zero modes, and the calculations done in the Appendices have included arbitrarily twisted boundary conditions to allow application to O($N)^\ast$ models.

\section*{Acknowledgements}
S.W. and S.S. thank S. Chatterjee and A. Thomson for numerous helpful conversations. M.S., L.-P.H. and A.M.L acknowledge support through the Austrian Science Fund SFB FoQus (F-4018).
The computational results presented have been achieved in part using the Vienna Scientific Cluster (VSC)
and the HPC infrastructure LEO of the University of Innsbruck. 
S.W. and S.S. were supported by the NSF under Grant DMR-1360789, and MURI grant W911NF-14-1-0003 from ARO. Research at Perimeter Institute is supported by the Government of Canada through Industry Canada and by the Province of Ontario through the Ministry of Research and Innovation. SS also acknowledges support from Cenovus Energy at Perimeter Institute.

\appendix
\section{Infinite volume computations}
\label{sec:infvol}

Here we recall some properties of the Wilson-Fisher fixed point in an infinite volume. We will need the resulting expressions to renormalize the theory in a finite volume, and we relate the couplings to universal observables of the infinite-volume theory. The computations are standard \cite{ZJ02}, so we will be brief. Here we express the theory as a Euclidean action,
\beq
\mathcal{S} = \int d\tau d^dx \left[ \frac{1}{2} \left( \partial_{\tau} \phi_{\alpha} \right)^2 + \frac{1}{2}\left( \nabla \phi_{\alpha} \right)^2 + \frac{s_0}{2} \phi_{\alpha}^2 + \frac{u}{4!} \phi_{\alpha}^4 + \Lambda \right]
\eeq
where $d = 3 - \epsilon$, but now we integrate over infinite volume. At one-loop, we need to renormalize the couplings $s$, $u$, and the energy density $\Lambda$ (there is no field renormalization until two-loop). To this end, we introduce the renormalized couplings
\bea
s_0 &=& s_c + Z_2 s \nn
u &=& Z_4 \frac{\mu^{\epsilon} g}{S_{d+1}}
\eea
where $S_d = 2/(\Gamma(d/2)(4\pi)^{d/2})$ is a convenient factor, and $\mu$ is an arbitrary renormalization scale. The renormalized coupling $s$ has been defined so that $s=0$ at criticality by definition. We renormalize the theory using a modified minimal subtraction scheme as detailed in Ref.~\cite{ZJ02}, where the renormalization constants are given by
\bea
Z_2 &=& 1 + \frac{N+2}{6 \epsilon}g \nn
Z_4 &=& 1 + \frac{N+8}{6 \epsilon} g
\label{z24}
\eea
and the critical coupling is $s_c = 0$. The Wilson-Fisher fixed point is obtained when the couplings take the values
\bea
s &=& 0 \nn
g^{\ast} &=& \frac{6 \epsilon}{N+8} \left( 1 + \frac{3(3N+14)}{(N+8)^2} \epsilon \right)
\eea

We are also interested in the vacuum energy in the vicinity of the Wilson-Fisher fixed point. To this end, we have included the additive constant $\Lambda$ which we choose to make the ground state energy density finite in an infinite volume. This constant also depends on the renormalization scale, and is given by
\bea
\Lambda &=& \frac{s^2 Z_2^2 S_{d+1}}{\mu^{\epsilon}} Z_{\Lambda} \nn
Z_{\Lambda} &=& \frac{N}{4 \epsilon} - \frac{N(N+2)}{24 \epsilon^2}g
\label{vacrenorm}
\eea
Our choice of regularization leads to the vanishing of the ground state energy density at $L = \infty$ and $s = s_c$, where the system has full conformal invariance.

We note that computations will involve the arbitrary energy scale $\mu$. This dependence can always be eliminated in favor of physical quantities. As an example, we can consider the exact energy gap in an infinite volume when $s>0$. The inverse propagator can be written as a function of the Euclidean momentum $p = (\omega, k)$ as
\beq
G(p)^{-1} = p^2 + s - \Sigma(p^2 )
\eeq
Then the exact energy gap $m$ is given by
\beq
m^2 = s - \Sigma(m^2)
\eeq
At one loop this gives
\beq
m^2 = s \left[ 1 + \frac{\epsilon}{2} \left(\frac{N+2}{N+8}\right) \ln (s/\mu^2) \right]
\eeq
where we have taken $g=g^{\ast}$. We can then always use this to rewrite dependence on $s$ and $\mu$ in terms of the physical parameter $m$ by inverting this expression,
\beq
s = m^2 \left[ 1 - \frac{\epsilon}{2}\left(\frac{N+2}{N+8}\right) \ln (m^2/\mu^2) \right]
\label{massdef}
\eeq

%For $s<0$, we could instead rewrite our results in terms of the characteristic energy scale in the ordered state, corresponding to the energy gap for $N=1$ and the spin stiffness for $N>1$. Instead we take a simpler approach of defining the energy scale for $s < 0$ in terms of the corresponding energy scale of its partner point in the disordered phase at $-s$. Then we can give our results in terms of the energy scale $m$, where $m$ is the physical energy gap for $s>0$, while $m$ is the physical energy gap at the corresponding point $-s$ for $s<0$. Since Eq.~(\ref{massdef}) is an even function of $m$, we simply use it to redefine $s$ in all expressions.
In an infinite volume, we can calculate the vacuum energy density. Setting $g = g^{\ast}$, this is given by
\bea
\frac{E_0}{\mathcal{A}^{d/2}} &=& \frac{N s^2}{64 \pi^2 \mu^{\epsilon}} \left[ \ln \left( s/\mu^2 \right) - \frac{1}{2}  \right] + \frac{N s^2 \epsilon}{128 \pi^2 \mu^{\epsilon}} \left[ 1 - E_{\gamma} + \ln 4 \pi \right] \left[ \ln \left( s/\mu^2 \right) - \frac{1}{2}  \right] \nn
&-& \frac{N s^2 \epsilon}{256 \pi^2 \mu^{\epsilon}} \left[ \ln^2 \left( s/\mu^2 \right) - \ln \left( s/\mu^2 \right) + \frac{\pi^2}{3} + \frac{1}{2} \right] + \frac{N(N+2)}{(N+8)}\frac{s^2 \epsilon}{\mu^{\epsilon}}\left[ \ln \left( s/\mu^2 \right) + \frac{\pi^2}{3} \right]
\eea
where $E_{\gamma}$ is the Euler-Mascheroni constant and $s>0$. 

We note that our finite-volume calculations must be analytic through the critical point $s = 0$, and the system remains disordered for $s<0$. Therefore, any singularities or branch cuts present in these expressions must cancel out in final results. Therefore we prefer to give our expressions in terms of $s$ and $\mu$ rather than $m$ and $E_0$, which are both non-analytic for $s=0$, and undefined for $s<0$.

\section{Derivation of the Bloch effective Hamiltonian}
\label{blochpert}
Here, we give a derivation of Eq.~(\ref{eqn:effham}), the effective Hamiltonian from Bloch's perturbation theory \cite{B58}. For the relation between this approach and other effective Hamiltonian approaches to perturbation theory, we direct the reader to Ref.~\cite{K74}. 

We begin by considering a degenerate subspace of the unperturbed spectrum, $\Omega_0 = \mathrm{Span}\{|\epsilon_0 \rangle \}$, where the states $|\epsilon_0 \rangle$ satisfy
\beq
H_0 |\epsilon_0 \rangle = \epsilon_0 |\epsilon_0 \rangle
\eeq
These states are split into distinct energies by the exact Hamiltonian $H = H_0 + V$,
\beq
H |\alpha \rangle = E_{\alpha} | \alpha \rangle, \qquad E_{\alpha} = \epsilon_0 + \mathcal{O}(V)
\eeq
Here, we have defined the orthonormal basis of states $|\alpha\rangle$ which diagonalize the exact Hamiltonian and reduce to the degenerate manifold $| \epsilon_0 \rangle$ in the $V = 0$ limit. We will define the space spanned by these states by $\Omega = \mathrm{Span}\{|\alpha \rangle \}$. If the perturbation is small, there should be a one-to-one correspondence between the spaces $\Omega_0$ and $\Omega$. We will assume the latter fact in what follows.

Let $P_0$ be the projection operator onto the space $\Omega_0$, and define the states
\beq
|\alpha_0 \rangle = P_0 | \alpha \rangle
\label{eqn:a0def}
\eeq
The set of states $| \alpha_0 \rangle$ are a particular basis spanning $\Omega_0$, although this basis is not orthonormal in general. We now claim that we can define a linear operator $U$ such that
\bea
&&U | \alpha_0 \rangle = | \alpha \rangle \nonumber \\
&&U | \phi \rangle = 0 \quad \forall | \phi \rangle \notin \Omega_0
\eea
Once the operator $U$ is found, we can construct an effective Hamiltonian which acts on the unperturbed subspace but gives the exact energy spectrum,
\bea
H_{\mathrm{eff}} = P_0 H U \nn
H_{\mathrm{eff}} |\alpha_0 \rangle = E_{\alpha} |\alpha_0 \rangle \label{effdef}
\eea
This Hamiltonian acts on the unperturbed subspace, but gives the exact spectrum of the interacting Hamiltonian. We can also obtain the exact eigenstates by $|\alpha \rangle = U |\alpha_0 \rangle$.

We now make a few clarifying comments on the above operators and states. We note that $\Omega_0$ and $\Omega$ represent subspaces within the same Hilbert space, and that these subspaces overlap by assumption. Since the linearly independent basis $| \alpha_0 \rangle$ defined via (\ref{eqn:a0def}) is not necessarily orthogonal, the operator $U$ is not unitary, so the standard intuition on changes of basis does not apply here. For example, since we can decompose any state $|\phi \rangle$ in the Hilbert space as
\beq
| \phi \rangle = P_0 | \phi \rangle  + \left( 1 - P_0 \right)| \phi \rangle 
\eeq
it follows that
\beq
U | \phi \rangle = U P_0|\phi \rangle
\eeq
so
\beq
U P_0 = U
\eeq
is an exact operator identity. In particular, this implies $U | \alpha \rangle = | \alpha \rangle$. Furthermore, the effective Hamiltonian defined in Eq.~(\ref{effdef}) is not necessarily Hermitian. Thus, even though the vectors $|\alpha_0 \rangle$ are eigenvectors of $H_{\mathrm{eff}}$, it does not follow that the effective Hamiltonian is diagonal in the $|\alpha_0 \rangle$ basis. This can also be inferred from the fact that the $|\alpha_0\rangle$ may not be orthogonal. Below, we will define a similarity transform which does allow the definition of a Hermitian effective Hamiltonian from Eq.~(\ref{effdef}).

We now find an explicit expression for $U$ for a completely general Hamiltonian. Starting from Schr\"odinger's equation,
\bea
H | \alpha \rangle &=& E_\alpha |\alpha \rangle \nn
\Rightarrow U P_0 H | \alpha \rangle &=& E_\alpha |\alpha \rangle.
\eea
Since $UP_0 = U$,
\beq
\left( H - U H \right) | \alpha \rangle = 0
\eeq
for all $|\alpha \rangle \in \Omega$. Furthermore, for any state $|\alpha \rangle \in \Omega$, we have $U | \alpha \rangle = | \alpha \rangle$ and $U H_0 |\alpha \rangle = \epsilon_0 | \alpha \rangle$, from which we have the general operator relation on the entire Hilbert space:
\bea
\left( H_0 + V - \epsilon_0 - UV \right) U = 0 \nn
\Rightarrow \left( \epsilon_0 - H_0 \right) U = VU - UVU
\eea
We can invert the left-hand side of this equation by using the fact that the operator $\left( \epsilon_0 - H_0 \right)$ satisfies $\left( \epsilon_0 - H_0 \right)P_0 = 0$ and has a well-defined inverse in the space orthogonal to $\Omega_0$. This gives the implicit equation
\beq
U = P_0 + \frac{1 - P_0}{\epsilon_0 - H_0} \left( VU - UVU \right)
\eeq
This allows an expansion in powers of $V$. Up to third order in $V$, this is given by
\bea
H_{\mathrm{eff}} &=& \epsilon_0 P_0 + P_0 V P_0 + P_0 V \frac{1 - P_0}{\epsilon_0 - H_0} V P_0 \nn
&+& P_0 V \frac{1 - P_0}{\epsilon_0 - H_0} V \frac{1 - P_0}{\epsilon_0 - H_0} V P_0 - P_0 V \frac{1 - P_0}{\left(\epsilon_0 - H_0 \right)^2} V P_0 V P_0 + \cdots
\label{eqn:effham2}
\eea
For higher order expressions, see Refs. \cite{B58,L83}. As seen in Section \ref{effhamcalcs}, the leading non-trivial two-loop diagram is of order $\epsilon^{5/3}$. We can now check explicitly by putting $V$ from Eq.~(\ref{intH}) that the last term in Eq.~(\ref{eqn:effham2}) is of order $\epsilon^{5/3}$ or higher, justifying our truncation of the effective Hamiltonian in the main body of our paper.

We note that the effective Hamiltonian defined above will not be Hermitian in general without a redefinition, although we will not encounter this problem at one-loop. In general, we have
\beq
B H_{\mathrm{eff}}^{\dagger} = H_{\mathrm{eff}} B, \qquad B \equiv P_0 P P_0
\eeq
where $P$ is the projection operator onto the space $\Omega$. Then the redefined Hermitian Hamiltonian
\beq
H'_{\mathrm{eff}} \equiv B^{-1/2} H_{\mathrm{eff}} B^{1/2}
\eeq
acts on $\Omega_0$ and has the same spectrum as $H_{\mathrm{eff}}$.

\section{Loops sums}
\label{dimreg}

In this appendix we give the calculation of the relevant loop diagrams on a torus in fractional dimensions. The torus is parametrized by complex coordinates, $x = x_1 + i x_2$, with two complex periods $\omega_1$ and $\omega_2$, see Fig.~\ref{fig:torus}. We define the modular parameter, $\tau = \omega_2/\omega_1 = \tau_1 + i \tau_2$, and the length scale $L \equiv |\omega_1|$. The area of the torus is given by $\mathcal{A} = \tau_2 L^2$.

In this geometry, the basis vectors of the dual lattice are
\beq
k_1 = -i\omega_2/\mathcal{A}, \qquad k_2 = i \omega_1/\mathcal{A}
\eeq
We consider the eigenvalues of the Laplacian on the torus. In the interest of generality, we will allow twisted boundary conditions along the two cycles of the torus. 
\bea
\phi_{\alpha}(x + \omega_1) = e^{2 \pi i a_1} \phi_{\alpha}(x) \nn
\phi_{\alpha}(x + \omega_2) = e^{2 \pi i a_2} \phi_{\alpha}(x)
\eea
The parameters $a_1$ and $a_2$ take values in the range $[0,1)$. For $a_{1,2}$ not equal to either an $1$ or $1/2$, the fields $\phi_{\alpha}$ are actually complex and our symmetry breaks down from O($N$) to $SU(N)\times U(1)$. In the main text of this paper we always take $a_1 = a_2 = 0$, but the calculation of loop sums on the torus with twisted boundary conditions has found applications in related work \cite{HLSSW16,WS16,WW-KS16}, so we give general results here.

The eigenvalues of the Laplacian are
\beq
|k_{n,m}|^2 = \left( 2 \pi \right)^2 \left| (n + a_1) k_1 + (m + a_2) k_2 \right|^2, \qquad n, m \in \mathbb{Z}
\eeq
and a general one-loop diagram will take the form
\beq
\sideset{}{'}\sum_{n,m \in \mathbb{Z}} \frac{1}{\left( |k_{n,m}|^2 + s \right)^\nu} = \left( \frac{ \tau_2 L}{2 \pi} \right)^{2\nu} \sideset{}{'}\sum_{n,m \in \mathbb{Z}} \frac{1}{\left( |m + a_2 + (n + a_1)\tau |^2 + \gamma \right)^\nu} \label{1loop}
\eeq
where $\gamma = \tau_2^2 L^2 s/4 \pi^2$, and the primed summation indicates that the $n = m = 0$ term is omitted in the fully periodic case $(a_1,a_2) = (0,0)$. 

We now generalize this sum to arbitrary dimension. This is done by promoting the two-vector $(n + a_1,m + a_2)$ to a $d$-dimensional vector where the first $d/2$ components are $n + a_1$ and the last $d/2$ components are $m + a_2$. Then in (\ref{1loop}) the sums are taken over $n, m \in \mathbb{Z}^{d/2}$. We write the sums as
\beq
S_{\nu}^{(d)}(s,\tau) = \sideset{}{'}\sum_{n,m \in \mathbb{Z}^{d/2}} \frac{1}{\left( |m + a_2 + (n + a_1)\tau |^2 + \gamma \right)^\nu}. \label{gdef}
\eeq

The summand is rewritten using the identity
\beq
\frac{1}{A^\nu} = \frac{\pi^\nu}{\Gamma(\nu)} \int_0^{\infty} d\lambda \lambda^{\nu-1} e^{- \pi \lambda A}
\eeq
giving
\beq
S_{\nu}^{(d)} = \frac{\pi^\nu}{\Gamma(\nu)} \int_0^{\infty} d\lambda \lambda^{\nu-1} e^{- \pi \lambda \gamma} \sideset{}{'}\sum_{n,m \in \mathbb{Z}^{d/2}} \exp \left(- \pi \lambda |m + a_2 + (n + a_1)\tau |^2\right).
\eeq
We can now write the sum in terms of the two-dimensional Riemann theta function, defined as
\beq
\Theta\left(\lambda,\mathbf{\Omega},\mathbf{u}\right) \equiv \sum_{\mathbf{n} \in \mathbb{Z}^2} \exp\big(-\pi \lambda \mathbf{n}^{\intercal} \cdot \mathbf{\Omega} \cdot \mathbf{n} - 2 \pi \mathbf{n}^{T} \cdot \mathbf{u} \big) \label{rtheta}
\eeq
where $\Omega$ is a $2\times 2$ matrix and $\mathbf{u}$ is a two-dimensional vector. Then
\beq
S_{\nu}^{(d)} = \frac{\pi^\nu}{\Gamma(\nu)} \int_0^{\infty} d\lambda \lambda^{\nu-1} e^{-\pi \lambda \gamma} \left[ \exp\left( - \frac{d \pi \lambda \eta}{2} \right) \Theta\left(\lambda,\mathbf{\Omega}(\tau),\mathbf{v}_1\right)^{d/2} - \delta_{a_1 0}\delta_{a_2 0} \right] \label{unreg}
\eeq
where
\beq
\mathbf{\Omega}(\tau) = 
\begin{pmatrix}
| \tau |^2 & \tau_1 \\
\tau_1 & 1 
\end{pmatrix}, \qquad \mathbf{v}_1 = \lambda \begin{pmatrix} \tau_1\left( a_2 + a_1 \tau_1 \right) + a_1 \tau_2^2 \\ a_2 + a_1 \tau_1 \end{pmatrix}, \qquad \eta = (a_1 \tau_2)^2 + (a_2 + a_1 \tau_1)^2.
\eeq

The function $S_{\nu}^{(d)}(s,\tau)$ has a divergence for small $\lambda$ whenever $\nu < d/2$. We evaluate the sums by dimensional regularization: we separate out the divergent parts and and evaluate them in the convergent regime $\nu > d/2$, and then analytically continue them to the physical values of $\nu$ and $d$ of interest, taking care to renormalize any poles in $\epsilon$ which arise. For our purposes, it is also crucial that we obtain final expressions which are regular at $s = 0$ and remain finite for $s < 0$, since the finite-volume theory should be analytic through the critical point.

We proceed by splitting the integral in Eq~(\ref{unreg}) into two parts, $\int_0^{\infty} = \int_0^1 + \int_1^{\infty}$, and studying the divergent part. Using the mathematical identity
\beq
\Theta\left(\lambda,\mathbf{\Omega},\mathbf{u}\right) = \frac{1}{\lambda \sqrt{\det \mathbf{\Omega}}} \exp \left( \frac{\pi}{\lambda} \mathbf{u}^T \cdot \mathbf{\Omega}^{-1} \cdot \mathbf{u} \right) \Theta\left(\frac{1}{\lambda},\mathbf{\Omega}^{-1},-\frac{i}{\lambda}\mathbf{\Omega}^{-1} \cdot\mathbf{u} \right),
\eeq
we write the lower portion of the integral as
\bea
&&\frac{\pi^\nu}{\Gamma(\nu)} \int_0^{1} d\lambda \lambda^{\nu-1} e^{- \pi \lambda \gamma} \left[ \tau_2^{-d/2} \lambda^{-d/2} \Theta\left(\frac{1}{\lambda},\mathbf{\Omega}(\tau)^{-1},\mathbf{v}_2\right)^{d/2} - \delta_{a_1 0}\delta_{a_2 0} \right] \nn
& =& \tau_2^{-d/2}\frac{\pi^\nu}{\Gamma(\nu)} \int_1^{\infty} d\lambda \lambda^{d/2-\nu-1} \left[e^{- \pi \gamma/\lambda} \Theta\left(\lambda,\mathbf{\Omega}(\tau)^{-1},\mathbf{v}_2\right)^{d/2} - 1 + \frac{\pi \gamma}{\lambda} - \frac{(\pi \gamma)^2}{2 \lambda^2} \right] \nn
&+& \ \tau_2^{-d/2}\frac{\pi^\nu}{\Gamma(\nu)} \int_1^{\infty} d\lambda \lambda^{d/2-\nu-1} \left( 1 - \frac{\pi \gamma}{\lambda} + \frac{(\pi \gamma)^2}{2 \lambda^2} \right) - \frac{\pi^\nu}{\Gamma(\nu)}\delta_{a_1 0}\delta_{a_2 0} \int_1^{\infty} d\lambda \lambda^{-\nu-1} e^{-\pi \gamma/\nu} \qquad \label{divpart1}
\eea
where we define the vector $\mathbf{v}_2 = -i(a_1, a_2)$. For $d=3$ and $\nu \geq -1/2$, the UV divergences are entirely contained in the last line of Eq~(\ref{divpart1}). The very last term can be integrated in its convergent regime, obtaining
\beq
\frac{1}{\Gamma(\nu)} \int_1^{\infty} d\lambda \lambda^{-\nu-1} e^{-\pi \gamma/\nu} = e^{-\pi \gamma} \sum_{k = 0}^{\infty} \frac{(\pi \gamma)^k}{\Gamma(\nu + k + 1)} \label{perpart}
\eeq
This expression was obtained by evaluating the integral for $\nu > 0$ and $\gamma > 0$, obtaining an expression involving incomplete gamma functions, and then expressing these functions as a power series. The final expression is a single-valued analytic function for all $\gamma$ and $\nu$ with no singularities \cite{AS64}, and evaluating this series numerically is trivial. 

When $\nu > d/2$ we can evaluate
\beq
\int_1^{\infty} d\lambda \lambda^{d/2-\nu-1} \left( 1 - \frac{\pi \gamma}{\lambda} + \frac{(\pi \gamma)^2}{2\lambda^2} \right) = \frac{1}{\nu - d/2} - \frac{\pi \gamma}{1 + \nu - d/2} + \frac{(\pi \gamma)^2}{4 + 2 \nu - d}
\eeq
Since we are expanding around $d=3$, the three terms will contribute poles for $\nu = 3/2$, $\nu = 1/2$, and $\nu = -1/2$ respectively. We will see that these three poles are related to the renormalization of $u$, $s_0$, and $\mathcal{E}_0$.

To summarize our results so far, we have written the loop sum as
\bea
\sideset{}{'}\sum_{n,m \in \mathbb{Z}} \frac{1}{\left( |k_{n,m}|^2 + s \right)^\nu} = \left( \frac{ \tau_2 L}{2 \pi} \right)^{2\nu} S_{\nu}^{(d)}(s,\tau),
\eea
\bea
S_{\nu}^{(d)} &=& \frac{\pi^\nu}{\Gamma(\nu)} \int_1^{\infty} d\lambda \lambda^{\nu-1} e^{-\pi \lambda \gamma} \left[ \exp\left( - \frac{d \pi \lambda \eta}{2} \right) \Theta\left(\lambda,\mathbf{\Omega}(\tau),\mathbf{v}_1\right)^{d/2} - \delta_{a_1 0}\delta_{a_2 0} \right] \nn
&+& \ \tau_2^{-d/2}\frac{\pi^\nu}{\Gamma(\nu)} \int_1^{\infty} d\lambda \lambda^{d/2-\nu-1} \left[ e^{- \pi \gamma/\lambda} \Theta\left(\lambda,\mathbf{\Omega}(\tau)^{-1},\mathbf{v}_2\right)^{d/2} - 1 + \frac{\pi \gamma}{\lambda} - \frac{(\pi \gamma)^2}{2 \lambda^2} \right] \nn 
&+& \tau_2^{-d/2}\frac{\pi^\nu}{\Gamma(\nu)} \left( \frac{1}{\nu - d/2} - \frac{\pi \gamma}{1 + \nu - d/2} + \frac{(\pi \gamma)^2}{4 + 2\nu - d} \right) - \delta_{a_1 0}\delta_{a_2 0} e^{-\pi \gamma} \sum_{k = 0}^{\infty} \frac{\pi^{k + \nu} \gamma^k}{\Gamma(\nu + k + 1)} \qquad \qquad \label{totsum}
\eea
where $\gamma = \tau_2^2 L^2 s/4 \pi^2$, the Riemann theta function $\Theta$ is defined in (\ref{rtheta}), and
\bea
&\mathbf{\Omega}(\tau) = 
\begin{pmatrix}
| \tau |^2 & \tau_1 \\
\tau_1 & 1 
\end{pmatrix}, \qquad
&\mathbf{\Omega}(\tau)^{-1} = 
\frac{1}{\tau_2^2}\begin{pmatrix}
1 & -\tau_1 \\
-\tau_1 & | \tau |^2
\end{pmatrix} \\
&\mathbf{v}_1 = \lambda \begin{pmatrix} \tau_1\left( a_2 + a_1 \tau_1 \right) + a_1 \tau_2^2 \\ a_2 + a_1 \tau_1 \end{pmatrix}, \qquad
&\mathbf{v}_2 = - i \begin{pmatrix} a_1 \\ a_2 
\end{pmatrix}, \qquad \eta = (a_1 \tau_2)^2 + (a_2 + a_1 \tau_1)^2.
\label{vardefns}
\eea

From Eq.~(\ref{totsum}) we define finite functions. They are given in terms of the $S_{\nu}^{(d)}$, but with their poles around $d=3$ removed. 
\bea
f_{-1/2}^{(3-\epsilon)}(\tau,s,\mu) &=& S_{-1/2}^{(3-\epsilon)} + \frac{\tau_2^{5/2} L^4 s^2}{4 \pi \mu^{\epsilon} \mathcal{A}^{\epsilon/2} \epsilon}S_{4 - \epsilon} \nn
f_{1/2}^{(3-\epsilon)}(\tau,s,\mu) &=& S_{1/2}^{(3-\epsilon)} + \frac{4 \pi \tau_2^{1/2} L^2 s}{\mu^{\epsilon} \mathcal{A}^{\epsilon/2} \epsilon}S_{4 - \epsilon} \nn
\tilde{f}_{3/2}^{(3-\epsilon)}(\tau,s,\mu) &=& S_{3/2}^{(3-\epsilon)} - \frac{32 \pi^3 \tau_2^{-3/2}}{\mu^{\epsilon} \mathcal{A}^{\epsilon/2} \epsilon}S_{4 - \epsilon}
\label{fdefs}
\eea
The three functions $f_{\nu}^{(d)}$ have a regular power series about $s = 0$ and $\epsilon = 0$, and the extra factors are defined to simplify expressions in our renormalization scheme. We note that these functions are all dependent on $\mu$, but the first two functions are independent of $\mu$ at $s = 0$, while the third function will be exchanged for a $\mu$-independent function below. We note the identities
\bea
\frac{\partial}{\partial s} f^{(d)}_{\nu}(\tau,s,\mu) &=& - \frac{\nu \tau_2^2 L^2}{4 \pi^2} f^{(d)}_{\nu + 1}(\tau,s,\mu) \nn 
f^{(d)}_{\nu}(\tau + 1,0,\mu) &=& f^{(d)}_{\nu}(\tau,0,\mu) \nn
\tau_2^{\nu} f^{(d)}_{\nu}(\tau,0,\mu) &=& \frac{\tau_2^{\nu}}{|\tau|^{2\nu}} f^{(d)}_{\nu}(1/\tau,0,\mu)
\label{fprops}
\eea
The first identity will be useful for renormalizing the bare mass, while the other two are a consequence of modular invariance. 

\section{Renormalization of the effective Hamiltonian}
\label{app:renorm}

Here we consider the renormalization of physical quantities. To this end, we calculate the effective Hamiltonian for the splitting of the Fock vacuum, $h_{k = 0}$, which contains most of the divergences which need to be considered.

We calculate each individual expression in Eq.~(\ref{unrenorm}). The constant term is
\bea
h^{(0)}_{k=0} &=& \mathcal{A}^{(3-\epsilon)/2} \Lambda + \frac{N}{2} \sum_{k \neq 0} \sqrt{|k|^2 + s_0} + \frac{u \mathcal{A}^{\epsilon/2}}{\mathcal{A}^{3/2}} \frac{N(N+2)}{4(4!)} \left[ \sum_{k \neq 0} \frac{1}{\sqrt{|k|^2 + s_0}} \right]^2 \nn 
&=& \mathcal{A}^{(3-\epsilon)/2}\Lambda + \frac{\pi N}{\tau_2 L} \left[ f_{-1/2}^{(3 - \epsilon)}(\tau,s Z_2,\mu) - \frac{\tau_2^{5/2} L^4 s^2}{4 \pi \mu^{\epsilon} \mathcal{A}^{\epsilon/2} \epsilon}S_{4 - \epsilon} \right] \nn
&+& \frac{g \mu^{\epsilon}}{S_{4 - \epsilon}}\frac{ \mathcal{A}^{\epsilon/2}}{ \tau_2^{3/2} L^3} \frac{N(N+2)}{4(4!)} \frac{\tau_2^2 L^2}{4 \pi^2} \left[ f_{1/2}^{(3-\epsilon)}(\tau,s,\mu) - \frac{4 \pi \tau_2^{1/2} L^2 s}{\mu^{\epsilon} \mathcal{A}^{\epsilon/2} \epsilon}S_{4 - \epsilon} \right]^2
\eea
By expanding out all of the terms, including the $g$-dependent $Z_2$ factor using Eq.~(\ref{z24}), and using the definition of $\Lambda$ from Eq.~(\ref{vacrenorm}), we find that all poles in $\epsilon$ cancel, as well as all factors of $S_{4 - \epsilon}$. Once the poles cancel, we set $g = g^{\ast}$, obtaining
\beq
h^{(0)}_{k = 0} = \frac{ \pi N}{\tau_2 L} f^{(3 - \epsilon)}_{-1/2}(\tau,s,\mu) + \frac{N (N+2)}{(N+8)}\frac{\epsilon}{8 L} \tau_2^{1/2} f^{(3)}_{1/2}(\tau,s,\mu)^2
\eeq

The coefficient of $\varphi_{\alpha}^2$ is
\bea
h^{(2)}_{k=0} &=& \frac{\varphi^2}{2}\tau_2^{1/2} L s Z_2  + \frac{u \mathcal{A}^{\epsilon/2}}{2 \tau_2 L^2 } \left( \frac{\delta_{\alpha \beta}}{12}\varphi^2 + \frac{1}{6} \varphi_{\alpha} \varphi_{\beta} \right) \sum_{\mathbf{k} \neq 0} \frac{\delta_{\alpha \beta}}{\sqrt{|k|^2 + s}} \nn
&=& \frac{\varphi^2}{2} \left\{ \tau_2^{1/2} L s \left( 1 + \frac{N+2}{6 \epsilon} g \right) + \left( \frac{N+2}{12} \right) \frac{g \mu^{\epsilon}\mathcal{A}^{\epsilon/2}}{\tau_2 L^2 S_{4 - \epsilon}} \left[ f_{1/2}^{(3-\epsilon)}(\tau,s,\mu) - \frac{4 \pi \tau_2^{1/2} L^2 s}{\mu^{\epsilon} \mathcal{A}^{\epsilon/2} \epsilon}S_{4 - \epsilon} \right] \right\} \nonumber
\eea
The poles in $\epsilon$ and factors of $S_{4-\epsilon}$ cancel, and after setting $g = g^{\ast}$ we obtain
\beq
h^{(2)}_{k=0} = \frac{1}{\sqrt{\tau_2} L} \frac{\varphi^2}{2} \left\{ \tau_2 L^2 s + 2 \pi \epsilon \left( \frac{N+2}{N+8} \right) \tau_2^{1/2} f^{(3)}_{1/2}(\tau,s,\mu) \right\}
\eeq

Finally, the divergence in the quartic term cancels similarly, and after setting $g = g^{\ast}$ we find
\bea
h^{(4)}_{k=0} &=& \frac{u \mathcal{A}^{\epsilon/2}}{\sqrt{\tau_2}L} \frac{\varphi^4}{4!} - \frac{\left( u \mathcal{A}^{\epsilon/2} \right)^2}{8 \tau_2^2 L^4} \left( \frac{\delta_{\alpha \beta}}{12}\varphi^2 + \frac{1}{6} \varphi_{\alpha} \varphi_{\beta} \right) \left( \frac{\delta_{\gamma \delta}}{12}\varphi^2 + \frac{1}{6} \varphi_{\gamma} \varphi_{\delta} \right) \sum_{k \neq 0} \frac{\delta_{\alpha \gamma}\delta_{\beta \delta} + \delta_{\alpha \delta}\delta_{\beta \gamma}}{\left( |k|^2 + s \right)^{3/2}} \nn
& \Rightarrow& \frac{1}{\sqrt{\tau_2} L} \frac{\varphi^4}{4!} \left( \frac{48 \pi^2 \epsilon}{N+8} \right) \left\{ \frac{\mu^{\epsilon} \mathcal{A}^{\epsilon/2}}{8 \pi^2 S_{4 - \epsilon}} - \frac{\tau_2^{3/2} \epsilon}{4 \pi}\tilde{f}^{(3)}_{3/2}(\tau ,s,\mu) + \frac{3 \left( 3N+14 \right)}{\left( N+8 \right)^2}\epsilon \right\}
\eea

We note all of these expressions appear to depend on the arbitrary scale $\mu$. This dependence actually drops out of $h^{(4)}_{k=0}$ to this order in $\epsilon$, and it drops out of all quantities at the critical point $s = 0$. We can write
\beq
\frac{\mu^{\epsilon} \mathcal{A}^{\epsilon/2}}{8 \pi^2 S_{4 - \epsilon}} - \frac{\tau_2^{3/2} \epsilon}{4 \pi}\tilde{f}^{(3)}_{3/2}(\tau ,s,\mu) = 1 - \frac{\tau_2^{3/2} \epsilon}{4 \pi}f^{(3)}_{3/2}(\tau ,s) + \mathcal{O}(\epsilon^2)
\eeq
where $f^{(3)}_{3/2}(\tau,s)$ is $\mu$-independent. The fact that all $\mu$ dependence vanishes at $g = g^{\ast}$ and $s = 0$ is a manifestation of the scale invariance of the critical theory. 

Combining the above results, the effective Hamiltonian takes the form given in Eq.~(\ref{zeromodeham}):
\begin{gather}
h_{k = 0} = \mathcal{E}_{k = 0} + \frac{1}{\sqrt{\tau_2} L}\left( \frac{\pi_{\alpha}^2}{2} + \frac{R}{2} \varphi_{\alpha}^2 + \frac{U}{4!} \left(\varphi_{\alpha}^2\right)^2 \right) \nn
\mathcal{E}_{k = 0} \equiv \frac{ \pi N}{\tau_2 L} f^{(3 - \epsilon)}_{-1/2}(\tau,s,\mu) + \frac{1}{\sqrt{\tau_2} L} \frac{N(N+2)}{N+8}\frac{\epsilon}{8}\tau_2 f_{1/2}^{(3)}(\tau,s,\mu)^2 \nn
R \equiv \tau_2 L^2 s + 2 \pi \epsilon \left( \frac{N+2}{N+8} \right) \tau_2^{1/2} f_{1/2}^{(3)}(\tau,s,\mu) \nn
U \equiv \frac{48 \pi^2 \epsilon}{N+8} \left\{ 1 - \frac{\tau_2^{3/2} \epsilon}{4 \pi}f_{3/2}^{(3)}(\tau,s) + \frac{3\left(3N + 14 \right)}{\left( N+8 \right)^2}\epsilon \right\} 
\end{gather}

To the order required, the special functions are
\bea
f_{-1/2}^{(3-\epsilon)} &=& -\frac{1}{2 \pi} \int_1^{\infty} d\lambda \lambda^{-3/2} \exp\left(-\frac{\lambda \tau_2^2 L^2 s}{4 \pi}\right) \left[ \exp\left( - \frac{\left(3-\epsilon\right) \pi \lambda \eta}{2} \right) \Theta\left(\lambda,\mathbf{\Omega}(\tau),\mathbf{v}_1\right)^{\left(3-\epsilon\right)/2} - \delta_{a_1 0}\delta_{a_2 0} \right] \nn
&-& \frac{\tau_2^{-\left(3-\epsilon\right)/2}}{2 \pi} \int_1^{\infty} d\lambda \lambda^{1-\epsilon/2} \left[ \exp\left(-\frac{\tau_2^2 L^2 s}{4 \pi \lambda}\right) \Theta\left(\lambda,\mathbf{\Omega}(\tau)^{-1},\mathbf{v}_2\right)^{\left(3-\epsilon\right)/2} - 1 + \frac{\tau_2^2 L^2 s}{4 \pi \lambda} - \frac{\tau_2^4 L^4 s^2}{32 \pi^2 \lambda^2} \right] \nn
&-& \frac{\tau_2^{-\left(3-\epsilon\right)/2}}{\pi} + \frac{\tau_2^{\left(1+\epsilon\right)/2} L^2 s}{8 \pi^2} - \frac{\delta_{a_1 0}\delta_{a_2 0}}{\sqrt{\pi}} \exp\left(-\frac{\lambda \tau_2^2 L^2 s}{4 \pi}\right) \sum_{k = 0}^{\infty} \frac{ \left(\tau_2^2 L^2 s/4 \pi \right)^k}{\Gamma(k + 1/2)} \nn
&+& \frac{\tau_2^{5/2}L^4 s^2}{64 \pi^3} \left[ 1 - E_{\gamma} - \ln \left( \frac{\tau_2^2 L^2 \mu^2}{4 \pi} \right) \right] + \frac{\tau_2^{5/2} L^4 s^2 \epsilon}{128 \pi^3} \Bigg\{ 1 + 2 \ln^2 2 + \ln 4 \pi - \frac{\pi^2}{12} \nn
&+& \frac{1}{2} E_{\gamma} \left[ E_{\gamma} - 2 (1 + \ln 4 \pi ) \right] + \frac{1}{2} \ln\pi \ln 16 \pi - \left[ 1 - E_{\gamma} - \ln \left( \frac{\sqrt{\tau_2} L \mu}{4 \pi} \right) \right] \ln \left( \sqrt{\tau_2} L \mu \right) \nn
&-& \frac{1}{2} \ln^2 \tau_2 \Bigg\} + \mathcal{O}(\epsilon^2)
\label{f-12}
\eea
\bea
f_{1/2}^{(3)} &=& \int_1^{\infty} d\lambda \lambda^{-1/2} \exp\left(-\frac{\lambda \tau_2^2 L^2 s}{4 \pi}\right) \left[ \exp\left( - \frac{3 \pi \lambda \eta}{2} \right) \Theta\left(\lambda,\mathbf{\Omega}(\tau),\mathbf{v}_1\right)^{3/2} - \delta_{a_1 0}\delta_{a_2 0} \right] \nn
&+& \tau_2^{-3/2} \int_1^{\infty} d\lambda \left[ \exp\left(-\frac{\tau_2^2 L^2 s}{4 \pi \lambda}\right) \Theta\left(\lambda,\mathbf{\Omega}(\tau)^{-1},\mathbf{v}_2\right)^{3/2} - 1 + \frac{\tau_2^2 L^2 s}{4 \pi \lambda} \right] - \tau_2^{-3/2} \nn 
&-& \delta_{a_1 0}\delta_{a_2 0} \sqrt{\pi} \exp\left(-\frac{\lambda \tau_2^2 L^2 s}{4 \pi}\right) \sum_{k = 0}^{\infty} \frac{ \left(\tau_2^2 L^2 s/4 \pi \right)^k}{\Gamma(k + 3/2)} + \frac{\sqrt{\tau_2} L^2 s}{4 \pi} \left[ 1 - E_{\gamma} - \ln \left( \frac{\tau_2^2 L^2 \mu^2}{4 \pi} \right) \right] \qquad
\label{f12}
\eea
\bea
f_{3/2}^{(3)} &=& 2 \pi \int_1^{\infty} d\lambda \lambda^{1/2} \exp\left(-\frac{\lambda \tau_2^2 L^2 s}{4 \pi}\right) \left[ \exp\left( - \frac{3 \pi \lambda \eta}{2} \right) \Theta\left(\lambda,\mathbf{\Omega}(\tau),\mathbf{v}_1\right)^{3/2} - \delta_{a_1 0}\delta_{a_2 0} \right]
\nn &+& 2 \pi \tau_2^{-d/2} \int_1^{\infty} d\lambda \lambda^{-1} \left[ \exp\left(-\frac{\tau_2^2 L^2 s}{4 \pi \lambda}\right) \Theta\left(\lambda,\mathbf{\Omega}(\tau)^{-1},\mathbf{v}_2\right)^{3/2} - 1\right] \nn 
&+& 2 \pi \tau_2^{-3/2} \ln \tau_2 - \delta_{a_1 0}\delta_{a_2 0} \pi^{3/2} \exp\left(-\frac{\lambda \tau_2^2 L^2 s}{4 \pi}\right) \sum_{k = 0}^{\infty} \frac{ \left(\tau_2^2 L^2 s/4 \pi \right)^k}{\Gamma(k + 5/2)}
\label{f32}
\eea
The function $f_{-1/2}^{(3-\epsilon)}$ should be expanded to first order in $\epsilon$. It is possible to exchange the parameters $s$ and $\mu$ for the infinite volume gap and ground state energies, but the latter are not analytic through the critical point so we keep the $\mu$ dependence in our final expressions.

\section{Correspondence with the large-$N$ expansion}
\label{app:largen}

In a recent paper, two of us have computed the spectrum of the O($N$) model at $N = \infty$ \cite{WS16}. In this approach, one starts from a saddle-point of the theory which already has an energy gap on the torus, so the theory is IR safe and the zero mode plays no special role in the expansion. In this Appendix we take the small-$\epsilon$ limit of our large-$N$ results and compare them to the large-$N$ limit of the $\epsilon$-expansion. We find exact agreement where possible in both methods. Here we limit ourselves to $s = 0$.

In the large-$N$ expansion, one begins by solving the gap equation on the torus, which can be written
\beq
\frac{1}{\mathcal{A}^{d/2}}\sum_{k} \frac{1}{\sqrt{k^2 + \Delta^2}} = \int \frac{d^d k}{(2 \pi)^d} \frac{1}{k}.
\eeq
We can evaluate this using the methods of Appendix \ref{dimreg}, where dimensional regularization sets the integral on the right-hand side to zero, obtaining from Eq.~(\ref{fdefs})
\beq
\frac{1}{\mathcal{A}^{(3-\epsilon)/2} \Delta} + \frac{\tau_2 L}{2 \pi \mathcal{A}^{(3-\epsilon)/2}} f_{1/2}^{(3 - \epsilon)}(\Delta^2,\tau,\mu) - \frac{\tau_2 L}{2 \pi \mathcal{A}^{(3-\epsilon)/2}} \frac{4 \pi \tau_2^{1/2} L^2 \Delta^2}{\mu^{\epsilon} \mathcal{A}^{\epsilon/2} \epsilon}S_{4 - \epsilon}  = 0.
\label{gapeqn}
\eeq
As $\epsilon \rightarrow 0$, the function $f^{(3)}_{1/2}$ is regular, while the last term in Eq.~(\ref{gapeqn}) diverges. This requires the first term to diverge in the same way, from which one already sees that to leading order $\Delta \sim \epsilon^{1/3}$. Continuing this process, one can explicitly solve Eq~(\ref{gapeqn}) perturbatively in $\epsilon$, making use of identities derived in Appendix \ref{dimreg}. The gap at $N=\infty$ is given by
\beq
\sqrt{\mathcal{A}} \Delta = \left( 4 \pi^2 \epsilon \right)^{1/3} + \frac{1}{3}\left( 2 \pi \right)^{1/3} \sqrt{\tau_2}f_{1/2}^{(3)}(0,\tau) \epsilon^{2/3} - \tau_2^{3/2}\frac{2 \left( f_{1/2}^{(3)}(0,\tau) \right)^3 + 27 f_{3/2}^{(3)}(0,\tau)}{162 \left( 2 \pi \right)^{1/3}} \epsilon^{4/3}+ \mathcal{O}\big(\epsilon^{5/3}\big)
\eeq
where the functions are identical to those defined in Eqs.~(\ref{f-12}-\ref{f32}), and the $\mu$-dependence has dropped out. 

Once the gap equation has been solved, the Hamiltonian of the theory at $N = \infty$ is given by
\beq
H = \mathcal{E}_0 + P\sum_{k} \sum_{\alpha} \sqrt{|k|^2 + \Delta^2} b^{\dagger}_{\alpha}(k) b_{\alpha}(k)P + \left(1-P\right) \sum_{n,k} E_{n}(k) b^{\dagger}_n(k) b_n(k) \left(1 - P \right).
\eeq
Here,
\beq
\mathcal{E}_0 = \frac{N}{2} \sum_k \sqrt{|k|^2 + \Delta^2}
\eeq
which should be evaluated order-by-order in $\epsilon$ using the techniques in Appendix \ref{dimreg}. The operator $P$ projects onto all states which contain no O($N$) singlets. Finally, the energies of the singlet states $E_{n}(k)$ are the solutions to the equation
\beq
\Pi(k,E_n(k)) = 0
\eeq
where
\beq
\Pi(k,\omega) = \frac{1}{\mathcal{A}^{(3-\epsilon)/2}} \sum_{\mathbf{q}} \frac{\sqrt{q^2 + \Delta^2} + \sqrt{(k+q)^2 + \Delta^2}}{2 \sqrt{(q^2 + \Delta^2)((k+q)^2 + \Delta^2)}((\sqrt{q^2 + \Delta^2} + \sqrt{(k+q)^2 + \Delta^2})^2 - \omega^2)}.
\eeq

We now consider a subset of the above spectrum; specifically, the states created by
\beq
H_{\mathrm{eff},k=0} = \mathcal{E}_0 + P \sum_{\alpha} \Delta b^{\dagger}_{\alpha}(0) b_{\alpha}(0)P + \left(1-P\right) E_{0}(0) b^{\dagger}_0(0) b_0(0) \left(1 - P \right)
\label{focklargen}
\eeq
This Hamiltonian creates two kinds of zero-momentum ``particles,'' with masses $\Delta$ and $E_{1}(0)$ respectively. The particle with mass $E_1(0)$ transforms as an O($N$) singlet, while the states with $\ell$ mass-$\Delta$ particles are in the $\ell$th symmetric traceless tensor representation of O($N$). We now argue that $H_{\mathrm{eff},k=0}$ is precisely the $N = \infty$ limit of the Fock vacuum Hamiltonian in the $\epsilon$-expansion.

We saw in Section \ref{sec:num} that the solutions to the Fock vacuum Hamiltonian satisfy the equation
\beq
\frac{1}{2} \left( - \frac{1}{\rho^{N-1}} \frac{\partial}{\partial r} \rho^{N-1} \frac{\partial}{\partial \rho} - \frac{\ell \left( \ell + N - 2 \right)}{\rho^{N-1}} + r \rho^2 + 2 \rho^4 \right) R_{n,\ell}(\rho) = E_{n,\ell} R_{n,\ell}(\rho)
\eeq
Recall that this describes states in the $\ell$th symmetric traceless tensor representation of O($N$). We now use the large-$N$ expansion in quantum mechanics. The idea is that the centrifugal term acts as an effective mass at large-$N$, resulting in a harmonic well at the stationary point of the effective radial potential provided it is well-behaved. For a review of this expansion, see Ref.~\cite{AC90}, which gives an explicit formula (3.2.20) for the spectrum to the first few orders in $1/N$. At $N=\infty$ we find the spectrum
\beq
E_{n,\ell} = \mathcal{E}_0 + \Delta \ell + E_0 n
\eeq
where $\mathcal{E}_0$ and $\Delta$ agree exactly with their expressions calculated in the large-$N$ expression Eq.~(\ref{focklargen}) to order $\epsilon^{4/3}$. Furthermore, while we cannot compare $E_0$ directly with the first zero of $\Pi(0,\omega)$, we have evaluated it numerically for small values of $\epsilon$ and found very good agreement. Finally, the irreducible representations of the states under the O($N$) symmetry agrees exactly.

We mention that the $1/N$ expansion in quantum mechanics is an easy way to obtain $1/N$ corrections to the Fock vacuum Hamiltonian compared to the field-theoretic methods in Ref.~\cite{WS16}. One may also attempt to calculate the spectrum for the $k>0$ effective Hamiltonians using the large-$N$ expansion, which has been successfully applied to non-isotropic Hamiltonians in atomic and molecular physics \cite{AC90}.

\section{Strong-coupling expansion of isotropic quartic oscillators}
\label{numerics}

In this appendix we give details of the numerical calculation of the spectrum of the isotropic quartic oscillator
\beq
\frac{1}{2} \left( - \frac{1}{\rho^{N-1}} \frac{\partial}{\partial r} \rho^{N-1} \frac{\partial}{\partial \rho} - \frac{\ell \left( \ell + N - 2 \right)}{\rho^{N-1}} + r \rho^2 + 2 \rho^4 \right) R_{n,\ell}(\rho) = E_{n,\ell} R_{n,\ell}(\rho)
\label{radeqn2}
\eeq
in the strong-coupling limit, finding the coefficients
\beq
E_{n,\ell} = \sum_{m=1}^{\infty} c_{n,\ell,m} r^m
\eeq
We tabulate the values of $c_{n,\ell,m}$ which we have calculated in Tables \ref{tab:n=2}-\ref{tab:n=4}.

We begin by solving Eq.~(\ref{radeqn2}) numerically for $r=0$. We first fix the asymptotic behavior by writing
\beq
R_{n,\ell}(\rho) = \psi_{n,\ell}(\rho) e^{-\sqrt{2}\rho^{3}/3}
\label{rewrite}
\eeq
The exponential factor takes into account the large-$\rho$ behavior implied by Eq.~(\ref{radeqn2}). We then use a shooting method, using known boundary conditions on the wave function at $\rho = 0$ (including an arbitrary normalization) to compute $\psi_{n,\ell}$ in Mathematica using DSolve for variable values of the energy until we identify an eigenstate.

Once we find the energy to sufficient accuracy, the function $\psi_{n,\ell}$ will not change much for smaller $\rho$, but will always blow up after some value of $\rho$. However, the actual wave function $R_{n,\ell}$ is exponentially suppressed, so we only need to obtain $\psi_{n,\ell}$ accurately for small values of $\rho$ to obtain an accurate wave function.

\begin{table}
\centering
\begin{tabular}{| c | c | c | c | c | c |}
\hline
 $N = 2$ & \multicolumn{5}{|c|}{$\ell$}  \\
 \hline
& 0 & 1 & 2 & 3 & 4  \\
\hline
$c_{0,\ell,0}$ & 1.47715 & 3.39815 & 5.65434 & 8.09067 & 10.7583  \\
\hline
$c_{0,\ell,1}$ & 0.258539 & 0.447039 & 0.605913 & 0.747439 & 0.877189  \\
\hline
$c_{0,\ell,2}$ & -0.012345 & -0.015633 & -0.017109 & -0.017109 & -0.018469  \\
\hline
$c_{0,\ell,3}$ & 0.000903 & 0.000806 & 0.000697 & 0.000613 & 0.000548  \\
\hline\hline
$c_{1,\ell,0}$ & 6.00339 & 8.70045 & 11.53475 & 14.50868 & 17.61616  \\
\hline
$c_{1,\ell,1}$ & 0.554312 & 0.682554 & 0.80824713 & 0.92837 & 1.04294  \\
\hline
$c_{1,\ell,2}$ & -0.011291 & -0.012457 & -0.013714 & -0.014705 & -0.015489  \\
\hline
$c_{1,\ell,3}$ & 0.000133 & 0.000236 & 0.000293 & 0.000315 & 0.000320  \\
\hline
\end{tabular}
\caption{The coefficients of the strong-coupling expansion for the two-dimensional quartic oscillator.}
\label{tab:n=2}
\end{table}

\begin{table}
\centering
\begin{tabular}{| c | c | c | c | c | c |}
\hline
 $N = 3$ & \multicolumn{5}{|c|}{$\ell$}  \\
 \hline
& 0 & 1 & 2 & 3 & 4  \\
\hline
$c_{0,\ell,0}$ & 2.393644 & 4.478039 & 6.830308 & 9.401160 & 12.159017  \\
\hline
$c_{0,\ell,1}$ & 0.357801 & 0.529165 & 0.678421 & 0.813557 & 0.938665  \\
\hline
$c_{0,\ell,2}$ & -0.014371 & -0.016492 & -0.017576 & -0.018230 & -0.018667  \\
\hline
$c_{0,\ell,3}$ & 0.000865 & 0.000749 & 0.000651 & 0.000578 & 0.000521  \\
\hline\hline
$c_{1,\ell,0}$ & 7.335730 & 10.099944 & 13.004563 & 16.046193 & 19.217579  \\
\hline
$c_{1,\ell,1}$ & 0.618248 & 0.746036 & 0.869032 & 0.986315 & 1.09832  \\
\hline
$c_{1,\ell,2}$ & -0.011790 & -0.013117 & -0.014238 & -0.015120 & -0.015816  \\
\hline
$c_{1,\ell,3}$ & 0.000188 & 0.000271 & 0.000307 & 0.000319 & 0.000320  \\
\hline
\end{tabular}
\caption{The coefficients of the strong-coupling expansion for the three-dimensional quartic oscillator.}
\label{tab:n=3}
\end{table}

\begin{table}
\centering
\begin{tabular}{| c | c | c | c | c | c |}
\hline
 $N = 4$ & \multicolumn{5}{|c|}{$\ell$}  \\
 \hline
& 0 & 1 & 2 & 3 & 4  \\
\hline
$c_{0,\ell,0}$ & 3.398150 & 5.624339 & 8.090668 & 10.758265 & 13.600878  \\
\hline
$c_{0,\ell,1}$ & 0.447038 & 0.605918 & 0.747451 & 0.877202 & 0.998248  \\
\hline
$c_{0,\ell,2}$ & -0.015634 & -0.017110 & -0.017939 & -0.018466 & -0.018830  \\
\hline
$c_{0,\ell,3}$ & 0.000806 & 0.000697 & 0.000612 & 0.000547 & 0.000497  \\
\hline\hline
$c_{1,\ell,0}$ & 8.700454 & 11.534729 & 14.508675 & 17.616152 & 20.849517  \\
\hline
$c_{1,\ell,1}$ & 0.682554 & 0.808247 & 0.928370 & 1.042942 & 1.152539  \\
\hline
$c_{1,\ell,2}$ & -0.012457 & -0.013714 & -0.014704 & -0.015489 & -0.016111  \\
\hline
$c_{1,\ell,3}$ & 0.000236 & 0.000293 & 0.000315 & 0.000320 & 0.000319  \\
\hline
\end{tabular}
\caption{The coefficients of the strong-coupling expansion for the four-dimensional quartic oscillator.}
\label{tab:n=4}
\end{table}

Once we obtain a numerically accurate energy and wave function for $r=0$, we use logarithmic perturbation theory to compute the expansion in $r$. This has the benefit of only needing the unperturbed energy and wave function, whereas the standard Rayleigh-Schr\"odinger expansion requires knowledge of many excited states to get accurate values for the coefficients. The starting point for logarithmic perturbation theory is to write the wave function as
\beq
R_{n,\ell} = \prod_{i=1}^n (\rho - \rho_i) e^{G(\rho)}
\eeq
where $\rho_i$ are the nodes of $R_{n,\ell}$, and then rewrite the eigenvalue equation as an equation for $G$. Then we assume an expansion in $r$:
\bea
G(\rho) &=& \sum_{m=0} G_m(\rho) r^m \nn
\rho_i &=& \sum_{m=0} \rho_{i,m} r^m \nn
E_{n,\ell} &=& \sum_{m=0} c_{n,\ell,m} r^m
\label{eq:logexp}
\eea
Inserting these definitions into our eigenvalue equation, the resulting differential equation is linear order-by-order in perturbation theory, so the energies are given in closed form in terms of integrals only involving the unperturbed functions $G_0(\rho)$, $E_{n,\ell,0}$, and $\rho_{i,0}$. For explicit details, we refer the reader to References \cite{AA79a,AA79b}. We give the coefficients we have calculated in Tables \ref{tab:n=2}-\ref{tab:n=4}.

\section{Numerical Simulations}
\label{app:ED}

To compute the critical torus energy spectrum numerically with exact
diagonalization (ED) we consider explicit lattice models with a critical point
known to be in the respective universality class. Typical Hamiltonians for such
models with critical points in the O(2) and O(3) universality classes were
given in Eqs.~\eqref{eq:O2_spin1}~-~\eqref{eq:O3_2D}.

In a first step, we calculate the low-energy spectrum of a given Hamiltonian on
finite periodic clusters with up to $N=36$ lattice sites at the respective
quantum critical point. The spectrum can be divided into $S^z$ symmetry sectors
combined with the irreducible representations of the lattice space-group
symmetry. The critical low-energy spectrum collapses as $1/\sqrt{N} = 1/L$ with
the linear system size $L$. We multiply the spectrum by $L$ to get rid of this
scaling. Finally, we extrapolate the scaled spectrum for the different system
sizes linearly in $1/N$ to the thermodynamic limit to obtain the universal
critical torus energy spectrum. This extrapolation approach was successfully
used and corroborated by large-scale quantum Monte Carlo simulations in
Ref.~\cite{HLSSW16} for the Ising CFT.

\begin{figure}[h]
 \centering
 \includegraphics[height=7cm]{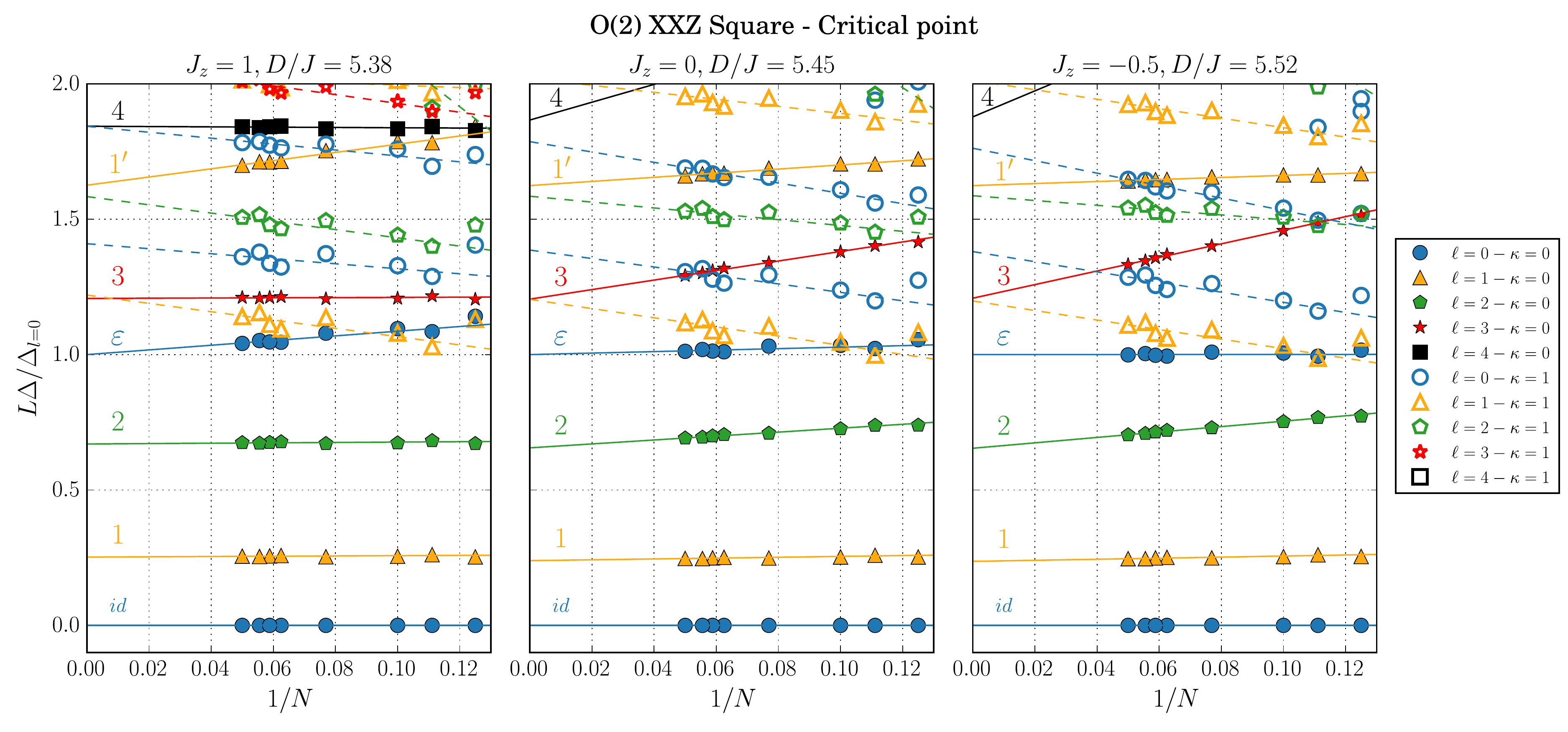}
 \caption{Finite-size critical low-energy spectrum for the Spin-1 O(2) model
     \eeqref{eq:O2_spin1} on a square lattice. The different panels show
     results for different couplings $J_z$. Filled symbols denote levels with
     zero momentum $\kappa=0$, empty symbols denote levels with $\kappa=1$. The
     different symbols encode different $\ell=S^z_{\text{tot}}$ sectors (see
     legend). The extrapolation to the thermodynamic limit is performed by
     linear fits in $1/N$. All levels are normalized by the gap
     $\Delta_{\ell=0}$. Although some levels on the finite-size systems for
     different $J_z$ are very different, the extrapolated levels are very close
     to each other. This strengthens the validity of our extrapolation approach
     and highlights the universal nature of the critical torus spectrum.}
     \label{fig:O2_XXZ}
\end{figure}

In Fig.~\ref{fig:O2_XXZ} we show the torus spectrum for the considered finite
square clusters with $N$ lattice sites for the model \eeqref{eq:O2_spin1}. The
different panels show results for different values of $J_z$.
After the linear extrapolation in $1/N$ to the thermodynamic limit the spectra
for the different $J_z$ agree to high precision with each other,
demonstrating its universal nature. 
As we cannot easily determine the speed of light for the models with ED, we
normalize all levels such that the extrapolated energy gap to the first excited
state with $\ell=0$ is set to one, $\Delta_{\ell=0} \equiv
\Delta_{\varepsilon_T} = 1$. After this, we have finally obtained the critical
torus energy spectrum from numerics, which is displayed in
Fig.~\ref{fig:O2_critical}. 

The same procedure is used to obtain the critical torus spectra for the other
Hamiltonians (not shown). There are, however, some subtleties to consider. First, for the
O(3) models $\ell = S_{\text{tot}}$ is usually not resolved in ED.
Nevertheless, $S_{\text{tot}}$ can usually be obtained by detecting levels with
the same energy with quantum numbers $-S_{\text{tot}} \leq S^z_{\text{tot}}
\leq S_{\text{tot}}$. 
Furthermore, for non-trivial magnetically ordered phases on one side of the
critical point, like the orderered state in model \eeqref{eq:O3_bilayer}, all
space-group symmetry sectors in the corresponding tower of states appear as
levels in the $\kappa=0$ sector. For the model \eeqref{eq:O3_bilayer} on the square
lattice this means that the levels in the $\kappa=0$ sector consist of levels with
momenta $\mathbf{k}=(0,0,0)$ and the Bragg peak momentum $\mathbf{k}=(0,0,\pi)$.
Subsequently, the $\kappa=1$ levels are composed of the momenta closest to $(0,0,0)$
as well as the momenta closest to $(0,0,\pi)$ etc. In QMC simulations the
antiferromagnetic version of this model with $J>0$ is usually considered. Then
the $\kappa=0$ levels would include levels with $\mathbf{k}=(0,0,0)$ and
$\mathbf{k}=(\pi,\pi,\pi)$.

In the following tables~\ref{tab:critO2EDsquare}~-~\ref{tab:critO3EDtriangular}
we list the numerical values for the critical torus spectra for the O(2) and
O(3) CFTs on both, square and triangular lattice geometries. Numerical values
for the Ising CFT can be found in the Supplemental Material of
Ref.~\cite{HLSSW16}.

\begin{table}
{%
\newcommand{\mc}[3]{\multicolumn{#1}{#2}{#3}}
\begin{center}
\begin{tabular}{|c||c|cccc||c|cccc}\hline%\cline{1-2}\cline{7-7}
$\tau=i$ & \mc{5}{c||}{$\kappa=0$} & \mc{5}{c|}{$\kappa=1$}\\\hline
 & $\ell=0$ & \mc{1}{c|}{$\ell=1$} & \mc{1}{c|}{$\ell=2$} &
\mc{1}{c|}{$\ell=3$} & $\ell=4$ & $\ell=0$ & \mc{1}{c|}{$\ell=1$} &
\mc{1}{c|}{$\ell=2$} & \mc{1}{c|}{$\ell=3$} & \mc{1}{c|}{$\ell=4$}\\\hline
× & 0 & \mc{1}{c|}{×} & \mc{1}{c|}{×} & \mc{1}{c|}{×} & × & × & \mc{1}{c|}{1.21} &
\mc{1}{c|}{} & \mc{1}{c|}{×} & \mc{1}{c|}{×}\\
× & × & \mc{1}{c|}{0.24} & \mc{1}{c|}{×} & \mc{1}{c|}{×} & × & 1.39 & \mc{1}{c|}{×}
& \mc{1}{c|}{×} & \mc{1}{c|}{×} & \mc{1}{c|}{×}\\
× & × & \mc{1}{c|}{×} & \mc{1}{c|}{0.66} & \mc{1}{c|}{×} & × & × & \mc{1}{c|}{×}
& \mc{1}{c|}{1.58} & \mc{1}{c|}{×} & \mc{1}{c|}{×}\\
× & 1.00 & \mc{1}{c|}{×} & \mc{1}{c|}{×} & \mc{1}{c|}{×} & × & 1.79 & \mc{1}{c|}{×}
& \mc{1}{c|}{×} & \mc{1}{c|}{×} & \mc{1}{c|}{×}\\
$L\Delta/\Delta_{\ell=0}$ & × & \mc{1}{c|}{×} & \mc{1}{c|}{×} & \mc{1}{c|}{1.21} & × & × & \mc{1}{c|}{2.02}
& \mc{1}{c|}{×} & \mc{1}{c|}{×} & \mc{1}{c|}{×}\\
× & × & \mc{1}{c|}{1.62} & \mc{1}{c|}{×} & \mc{1}{c|}{×} & × & × & \mc{1}{c|}{×}
& \mc{1}{c|}{×} & \mc{1}{c|}{2.12} & \mc{1}{c|}{×}\\
× & × & \mc{1}{c|}{×} & \mc{1}{c|}{×} & \mc{1}{c|}{×} & 1.86 & × & \mc{1}{c|}{×}
& \mc{1}{c|}{×} & \mc{1}{c|}{} & \mc{1}{c|}{2.78}\\
× & × & \mc{1}{c|}{×} & \mc{1}{c|}{2.40} & \mc{1}{c|}{×} & × & × & \mc{1}{c|}{×}
& \mc{1}{c|}{2.92} & \mc{1}{c|}{×} & \mc{1}{c|}{×}\\
 & × & \mc{1}{c|}{×} & \mc{1}{c|}{×} & \mc{1}{c|}{3.14} & ×
& × & \mc{1}{c|}{×} & \mc{1}{c|}{×} & \mc{1}{c|}{×} & \mc{1}{c|}{×}\\
× & × & \mc{1}{c|}{×} & \mc{1}{c|}{×} & \mc{1}{c|}{×} & 3.79 & × & \mc{1}{c|}{×}
& \mc{1}{c|}{×} & \mc{1}{c|}{×} & \mc{1}{c|}{×}\\
\hline
\end{tabular}
\end{center}
}%
\caption{Low-lying spectrum of the O(2) model from ED for geometry $\tau=i$. The
    given values are obtained by averaging over the results for the different
    considered models/parameters. Only the lowest levels in the fully symmetric
    representation regarding the point-group symmetry are listed.}
\label{tab:critO2EDsquare}
\end{table}

\begin{table}
{%
\newcommand{\mc}[3]{\multicolumn{#1}{#2}{#3}}
\begin{center}
\begin{tabular}{|c||c|cccc||c|cccc}\hline
$\tau=\frac{1}{2} + \frac{\sqrt{3}}{2}i$ & \mc{5}{c||}{$\kappa=0$} & \mc{5}{c|}{$\kappa=1$}\\\hline
& $\ell=0$ & \mc{1}{c|}{$\ell=1$} & \mc{1}{c|}{$\ell=2$} &
\mc{1}{c|}{$\ell=3$} & $\ell=4$ & $\ell=0$ & \mc{1}{c|}{$\ell=1$} &
\mc{1}{c|}{$\ell=2$} & \mc{1}{c|}{$\ell=3$} & \mc{1}{c|}{$\ell=4$}\\\hline
× & 0 & \mc{1}{c|}{×} & \mc{1}{c|}{×} & \mc{1}{c|}{×} & × & × & \mc{1}{c|}{1.26} &
\mc{1}{c|}{} & \mc{1}{c|}{×} & \mc{1}{c|}{×}\\
× & × & \mc{1}{c|}{0.25} & \mc{1}{c|}{×} & \mc{1}{c|}{×} & × & 1.46 & \mc{1}{c|}{×}
& \mc{1}{c|}{×} & \mc{1}{c|}{×} & \mc{1}{c|}{×}\\
× & × & \mc{1}{c|}{×} & \mc{1}{c|}{0.68} & \mc{1}{c|}{×} & × & × & \mc{1}{c|}{×}
& \mc{1}{c|}{1.65} & \mc{1}{c|}{×} & \mc{1}{c|}{×}\\
× & 1.00 & \mc{1}{c|}{×} & \mc{1}{c|}{×} & \mc{1}{c|}{×} & × & 1.83 & \mc{1}{c|}{×}
& \mc{1}{c|}{×} & \mc{1}{c|}{×} & \mc{1}{c|}{×}\\
$L\Delta/\Delta_{\ell=0}$ & × & \mc{1}{c|}{×} & \mc{1}{c|}{×} & \mc{1}{c|}{1.24} & × & × & \mc{1}{c|}{2.09}
& \mc{1}{c|}{×} & \mc{1}{c|}{×} & \mc{1}{c|}{×}\\
× & × & \mc{1}{c|}{1.64} & \mc{1}{c|}{×} & \mc{1}{c|}{×} & × & × & \mc{1}{c|}{×}
& \mc{1}{c|}{×} & \mc{1}{c|}{2.19} & \mc{1}{c|}{×}\\
× & × & \mc{1}{c|}{×} & \mc{1}{c|}{×} & \mc{1}{c|}{×} & 1.90 & × & \mc{1}{c|}{×}
& \mc{1}{c|}{×} & \mc{1}{c|}{} & \mc{1}{c|}{2.67}\\
× & × & \mc{1}{c|}{×} & \mc{1}{c|}{2.39} & \mc{1}{c|}{×} & × & × & \mc{1}{c|}{×}
& \mc{1}{c|}{2.85} & \mc{1}{c|}{×} & \mc{1}{c|}{×}\\
 & × & \mc{1}{c|}{×} & \mc{1}{c|}{×} & \mc{1}{c|}{3.23} & ×
& × & \mc{1}{c|}{×} & \mc{1}{c|}{×} & \mc{1}{c|}{×} & \mc{1}{c|}{×}\\
× & × & \mc{1}{c|}{×} & \mc{1}{c|}{×} & \mc{1}{c|}{×} & 3.84 & × & \mc{1}{c|}{×}
& \mc{1}{c|}{×} & \mc{1}{c|}{×} & \mc{1}{c|}{×}\\
\hline
\end{tabular}
\end{center}
}%
\caption{Low-lying spectrum of the O(2) model from ED for geometry
    $\tau=\frac{1}{2}+\frac{\sqrt{3}}{2}i$. The given values are obtained by
    averaging over the results for the different considered models/parameters.
    Only the lowest levels in the fully symmetric representation regarding the
point-group symmetry are listed.} 
\label{tab:critO2EDtriangular}
\end{table}

\begin{table}
{%
\begin{center}
\begin{tabular}{|c||c|c|c|c|c|}\hline
$\tau=i$ & $\ell=0$ & $\ell=1$ & $\ell=2$ &
$\ell=3$ & $\ell=4$ \\\hline
 & 0 &  &  &  &  \\
 &  & 0.29 &  &  & \\
 &  &  & 0.69 &  & \\
 & 1.00 &  &  &  & \\
 $L\Delta/\Delta_{\ell=0}$ &  &  &  & 1.19 & \\
 &  & 1.55 &  &  & \\
 &  &  &  &  & 1.76 \\
 &  &  & 2.17 &  & \\
 &  &  &  & 2.83 & \\
 &  &  &  &  & 3.28 \\
\hline
\end{tabular}
\end{center}
}%
\caption{Low-lying spectrum of the O(3) model from ED for geometry $\tau=i$.
    The given values are obtained from the results of the O(3) bilayer model.
    Only the lowest levels in the fully symmetric representation regarding the
    point-group symmetry are listed.}
\label{tab:critO3EDsquare}
\end{table}

\begin{table}
{%
\begin{center}
\begin{tabular}{|c||c|c|c|c|c|}\hline
$\tau=\frac{1}{2}+\frac{\sqrt{3}}{2}i$ & $\ell=0$ & $\ell=1$ & $\ell=2$ &
$\ell=3$ & $\ell=4$ \\\hline
 & 0 &  &  &  &  \\
 &  & 0.24 &  &  & \\
 &  &  & 0.60 &  & \\
 & 1.00 &  &  &  & \\
 $L\Delta/\Delta_{\ell=0}$ &  &  &  & 1.05 & \\
 &  & 1.49 &  &  & \\
 &  &  &  &  & 1.59 \\
 &  &  & 2.08 &  & \\
 &  &  &  & 3.01 & \\
 &  &  &  &  & 3.58 \\
\hline
\end{tabular}
\end{center}
}%
\caption{Low-lying spectrum of the O(3) model from ED for geometry
    $\tau=\frac{1}{2} + \frac{\sqrt{3}}{2}i$. The given values are obtained from
    the results of the O(3) bilayer model. Only the lowest levels in the fully
    symmetric representation regarding the point-group symmetry are listed.}
\label{tab:critO3EDtriangular}
\end{table}

\bibliography{epsilon-expansion}
\end{document}